\documentclass[12pt]{JHEP3}

\usepackage{epsfig}
\usepackage{epstopdf}
\usepackage{graphicx}

\title{Charged Lifshitz Black Holes}

\author{M. H. Dehghani\\
Physics Department and Biruni Observatory,
College of Sciences,\\ Shiraz University, Shiraz 71454, Iran\\
E-mail: \email{mhd@shirazu.ac.ir}}
\author{R. B. Mann\\ Department of Physics, University of Waterloo, 200 University
Avenue West, Waterloo, Ontario, Canada, N2L 3G1\\
E-Mail: \email{rbmann@sciborg.uwaterloo.ca}}
\author{R. Pourhasan\\ Department of Physics, University of Waterloo, 200 University
Avenue West, Waterloo, Ontario, Canada, N2L 3G1\\
E-Mail: \email{rpourhas@uwaterloo.ca}}

\abstract{
We investigate modifications of the Lifshitz black hole
solutions due to the presence of Maxwell charge  in higher
dimensions for arbitrary $z$ and any topology. We find that the
behaviour of large black holes is insensitive to the topology of
the solutions, whereas for small black holes significant
differences emerge. We generalize a relation previously obtained
for neutral Lifshitz black branes, and study more generally the
thermodynamic relationship between energy, entropy, and chemical
potential.  We also consider the effect of Maxwell charge on the
effective potential between objects in the dual theory. }

\begin{document}
\tableofcontents

\section{Introduction}

Since the AdS/CFT correspondence was proposed by Maldacena
\cite{Maldacena1} holography has been a useful tool in studying
strongly coupled field theories. Specifically, holography proposes
a duality between the gravitational dynamics in an asymptotically
AdS spacetime and a conformal field theory on the boundary.

In recent years the idea of holographic duality has been developed
beyond high energy physics to describe strongly coupled systems in
condensed matter physics, such as quantum critical systems
\cite{Hartnoll, Rokhsar, Ardonne, Vishwanath}. Such systems are
difficult to study using traditional methods in condensed matter based on weakly
interacting quasiparticles and broken symmetry.
Quantum critical points have a spacetime scale invariance which
provides a strong kinematic connection to the some versions of
AdS/CFT correspondence. This scaling symmetry is based on
anisotropic scaling transformation between space and time known as
Lifshitz scaling
\begin{equation}
t\rightarrow \lambda ^{z}t,\qquad r\rightarrow \lambda ^{-1}r,\qquad \mathbf{%
x}\rightarrow \lambda \mathbf{x},  \label{scalesym}
\end{equation}
where $z(\geq 1)$ is a dynamical critical exponent represents the
degree of anisotropy between space and time; manifestly $z=1$
exhibits relativistic systems. This scaling property (noted previously in other contexts \cite{Koroteev})
holographically is represented in the following form of the
spacetime metric \cite{Kachru}:
\begin{equation}
ds^{2}=\ell ^{2}\left( -r^{2z}dt^{2}+\frac{dr^{2}}{r^{2}}+r^{2}d\mathbf{x}%
^{2}\right) ,  \label{asmet}
\end{equation}
where the coordinates ($t,r,x^{i}$) are dimensionless and the only
length scale in the geometry is $\ell $. Metrics asymptotic to
(\ref{asmet}) can be generated as solutions to the equations of
motion that follow from the action:
\begin{equation}
I=\frac{1}{16\pi}\int d^{n+1}x\sqrt{-g}\left( R-2\Lambda
-\frac{1}{4}F_{\mu \nu }F^{\mu \nu }-\frac{1}{4}H_{\mu \nu }H^{\mu
\nu }-\frac{C}{2}B_{\mu }B^{\mu }\right) ,  \label{action}
\end{equation}
where $\Lambda $ is the cosmological constant, $F_{\mu \nu
}=\partial _{\lbrack \mu }A_{\nu ]}$ with $A_{\mu }$ representing
the Maxwell gauge and $H_{\mu \nu }=\partial _{\lbrack \mu }B_{\nu
]}$ is the field strength of the gauge field $B_{\mu }$ with mass
$m^{2}=C$.

Recently a similar action to (\ref{action}) but without a massless
gauge field (the Maxwell field) has been introduced in four
dimensions \cite{Daniel, Robb}, such that background metrics with
the anisotropic scale invariance (\ref{asmet}) are obtained as
exact solutions. An extension of these results to 5-dimensions
including the massless gauge field was carried out shortly
afterward \cite{Zingg0,Zingg}.  The introduction of the Maxwell
field introduced a new length scale, allowing the gravitational
system to undergo phase transitions. Working in (4+1) dimensions
\cite{Zingg0}, an exact charged solution for $z=6$ was obtained,
and the relationship between black hole temperature and charge was
numerically computed. Exact solutions in (4+1) dimensions
\cite{Zingg}, and then $(n+1)$ dimensions \cite{Zingg2,Pang} were
subsequently obtained, where $z=2n-2$.  These models yield a
holographic description of a strongly coupled quantum critical
point in $(n+1)$ dimensions with asymmetric scaling that models
the anomalous specific heat found at low temperature in many heavy
fermion compounds \cite{Zingg2}.

While some work has been done to obtain general numerical
solutions for charged Lifshitz black holes with planar topologies
\cite{Zingg0,Zingg,Zingg2}, the general character of such
solutions for general topologies and in arbitrary dimension has
not been explicated up to now. Here we investigate modifications
of the Lifshitz black hole solutions due to the presence of
Maxwell charge  in higher dimensions for arbitrary $z$. As with
their neutral counterparts, we find that the behaviour of large
black holes is insensitive to the topology of the solutions,
whereas for small black holes significant differences emerge.

We also investigate the thermodynamic relationship between energy,
entropy, and chemical potential, obtaining a generalization of a
relation previously obtained for neutral Lifshitz black holes
\cite{Peet,Peet2,HR}.

Accordingly, in Sec. 2 we introduce the field equation for the
metric functions and gauge field while our method is a slightly
different from what has been done in \cite{Zingg2}. In section 3
we discuss the behaviour of solutions at large $r$ and then go on
to examine near horizon expansions in section 4. Section 5 starts
by comparing numerical solutions in different dimensions for
uncharged black holes and follows by investigating the effect that
the Maxwell charge has on their behaviour. We consider the thermal
behaviour of these black holes and how charge affects their
temperature in section 6. We then go on to consider how charge
modifies the potential between two particles in the dual theory by
investigating corrections to the Wilson loop in 4-dimensions.  We
compute the conserved charge and the relationship between  energy,
entropy, and electromagnetic potential in section 8.  We close our
paper with some concluding remarks.

\section{Field Equations in ($n+1$)-dimensions}

Using the variational principle the field equations that follow
from the action (\ref{action}) are:
\begin{eqnarray}
&&G_{\mu\nu}+\Lambda g_{\mu\nu}=T_{\mu\nu},  \label{EqG} \\
&&\nabla^{\mu}H_{\mu\nu}=CB_{\mu},  \label{EqH} \\
&&\partial_{[\mu}B_{\nu]}=H_{\mu\nu},  \label{EqB} \\
&&\nabla^{\mu}F_{\mu\nu}=0,  \label{Eqk}
\end{eqnarray}
where the equations for the massive gauge field have been
rewritten in first-order form and where
\begin{equation}
T_{\mu\nu}=-\frac{1}{2}\left(\frac{1}{4}F_{\rho\sigma}F^{\rho\sigma}g_{\mu%
\nu}-F_{\phantom{\rho}{\mu}}^{\rho} F_{\rho\nu}+\frac{1}{4}%
H_{\rho\sigma}H^{\rho\sigma}g_{\mu\nu}-H_{\phantom{\rho}{\mu}}^{\rho}
H_{\rho\nu}+C\left[\frac{1}{2}B_{\rho}B^{\rho}g_{\mu\nu}-B_{\mu}B_{\nu}%
\right]\right)
\end{equation}
is the energy-momentum tensor of gauge fields.

The $(n+1)$-dimensional metric preserving the basic symmetries
(\ref {scalesym}) under consideration can be written as:
\begin{equation}
ds^{2}=\ell^2\left(-r^{2z}f^{2}(r)dt^{2}+\frac{g^{2}(r)dr^{2}}{r^{2}}%
+r^{2}d\Omega^{2}_{k}\right),  \label{metric}
\end{equation}
where
\begin{equation}
d\Omega^{2}_{k}=\left\{
\begin{array}{cc}
d\theta
_{1}^{2}+\sum\limits_{i=2}^{n-1}\prod\limits_{j=1}^{i-1}\sin
^{2}\theta _{j}d\theta _{i}^{2} & k=1 \\
d\theta _{1}^{2}+\sinh ^{2}\theta _{1}\left(d\theta
_{2}^{2}+\sum\limits_{i=3}^{n-1}\prod\limits_{j=2}^{i-1}\sin
^{2}\theta
_{j}d\theta _{i}^{2}\right) & k=-1 \\
\sum\limits_{i=1}^{n-1}d\theta _{i}^{2} & k=0
\end{array}
\right.
\end{equation}
represents the metric of an $(n-1)$-dimensional hypersurface with
constant curvature $(n-1)(n-2)k$ and volume $V_{n-1}$. The
hypersurface is $S^{n-1}$, $R^{n-1}$ or $H^{n-1}$, respectively,
for $k=1$, $0$ or $-1$.

The gauge fields are assumed to be
\begin{equation}
A_{t}=\ell r^{z}\kappa(r),\qquad B_{t}=q\ell r^{z}f(r)j(r),\qquad
H_{tr}=q\ell zr^{z-1}g(r)h(r)f(r), \label{gaugeans}
\end{equation}
with all other components either vanishing or being given by
antisymmetrization. In order to get the asymptotic Lifshitz
geometry (\ref
{asmet}) we demand $f(r)=g(r)=h(r)=j(r)=1$ and $\kappa(r)=0$ as $r\to \infty$%
, which in turn imposes the following constraints
\begin{eqnarray}
&&C=\frac{(n-1)z}{\ell^{2}},\qquad q^{2}=\frac{2(z-1)}{z},  \nonumber \\
&&\Lambda=-\frac{(z-1)^{2}+n(z-2)+n^{2}}{2\ell^{2}},  \label{cons}
\end{eqnarray}

Applying the ansatz (\ref{metric}) to the equation (\ref{Eqk})
yields:
\begin{equation}
(r^z\kappa)^{\prime}=\frac{Q}{r^{n-z}}fg \label{dk}
\end{equation}
where $Q$ is an  integration constant related to the Maxwell
charge (as we will discuss later) and we have chosen boundary
conditions such that the Maxwell vector potential vanishes at the
horizon. Substituting (\ref{cons}) and (\ref{dk}) into eqs.
(\ref{EqG}-\ref{EqB}), the field equations reduce to the system of
first order differential equations
\begin{eqnarray}
r\frac{df}{dr}&=&\frac{f}{4(n-1)r^{2}}\{2\left[%
(n-1)(z-1)j^{2}-z(z-1)h^{2}+(z-1)^{2}+n(z-2)+n^{2}\right]r^{2}g^{2}
\nonumber \\
&&+2(n-1)\left[(n-2)k\ell^{2}g^{2}-(n+2z-2)r^{2}\right]%
-Q^{2}r^{2(2-n)}g^{2}\},  \label{Eqf} \\
r\frac{dg}{dr}&=&\frac{g}{4(n-1)r^{2}}\{2\left[%
(n-1)(z-1)j^{2}+z(z-1)h^{2}-(z-1)^{2}-n(z-2)-n^{2}\right]r^{2}g^{2}
\nonumber \\
&&-2(n-1)\left[(n-2)k\ell^{2}g^{2}-nr^{2}\right]%
+Q^{2}r^{2(2-n)}g^{2}\},  \label{Eqg} \\
r\frac{dj}{dr}&=&-\frac{j}{4(n-1)r^{2}}\{2\left[%
(n-1)(z-1)j^{2}-z(z-1)h^{2}+(z-1)^{2}+n(z-2)+n^{2}\right]r^{2}g^{2}
\nonumber \\
&&+2(n-1)\left[(n-2)k\ell^{2}g^{2}-(n-2)r^{2}\right]%
-Q^{2}r^{2(2-n)}g^{2}\}+zgh,  \label{Eqj} \\
r\frac{dh}{dr}&=&(n-1)(jg-h),  \label{Eqh}
\end{eqnarray}
One can easily check that the above system will reduce to the four
dimensional Einstein case introduced in \cite{Daniel,Robb} if one
put $n=3$ and $z=2$ and $Q=0$.

\section{Exact Solutions}

In the previous section we introduced a set of four first order
differential equations which in general can not be solved
analytically because of non-linearity. However, it is possible to
find some exact solutions under certain assumptions. Indeed, if
one chooses the fields such that
\begin{equation}
f(r)=j(r)=\frac{1}{g(r)}\label{fgj}
\end{equation}
then the set of four ODE's reduces to three. One may easily solve
the differential equation for $h(r)$ which yields:
\begin{equation}
h(r)=1+\frac{\tilde{h}}{r^{n-1}},\label{EhA}
\end{equation}
where $\tilde{h}$ is an integration constant. Now, considering the
assumption (\ref{fgj}) and substituting eq. (\ref{EhA}), one may
find that Eqs. (\ref{Eqg}) and (\ref{Eqj}) would be satisfied
provided:
\begin{eqnarray}
2r^5(2-2n+z)\left[Q^2+2z(z-1)\tilde{h}^2\right]
+4zr^{n+4}(2z^2+2n-z(n+1)-2)\tilde{h}
\nonumber\\-4k\ell^{2}r^{2n+1}(n-1)(n-2)(z-2)=0.\label{EA}
\end{eqnarray}
There are two possibilities for solving this equation.

First, we can set $z=2(n-1)$. Then the two first terma in
(\ref{EA}) are eliminated. By solving the remaining terms for
$\tilde{h}$ we have
\begin{equation}
\tilde{h}=\frac{(n-2)^2k\ell^2r^{n-3}}{2(n-1)(3n-4)}.
\end{equation}
Since $\tilde{h}$ should be a constant then the solutions are
consistent if $k=0$ or $n=2,\,3$. The exact solutions for $k=0$
and arbitrary $n$ are given by \cite{ZinggQ}:
\begin{equation}
h=1, \quad f^2=j^2=\frac{1}{g^2}=1-\frac{Q^2}{2(n-1)^2r^{2n-2}},
\quad
\kappa=\frac{Q}{(n-1)r^{n-1}}\left(1-\frac{r^{n-1}_0}{r^{n-1}}\right)
\end{equation}
where there is just one horizon located at
\begin{equation}
r_{0}^{2}=\left[\frac{Q^2}{2(n-1)^2}\right]^{1/(n-1)},
\end{equation}
and the metric functions start from zero at the horizon and
monotonically increase to approach unity at infinity.

\begin{figure}[tbp]
\centering
{\includegraphics[width=.4\textwidth]{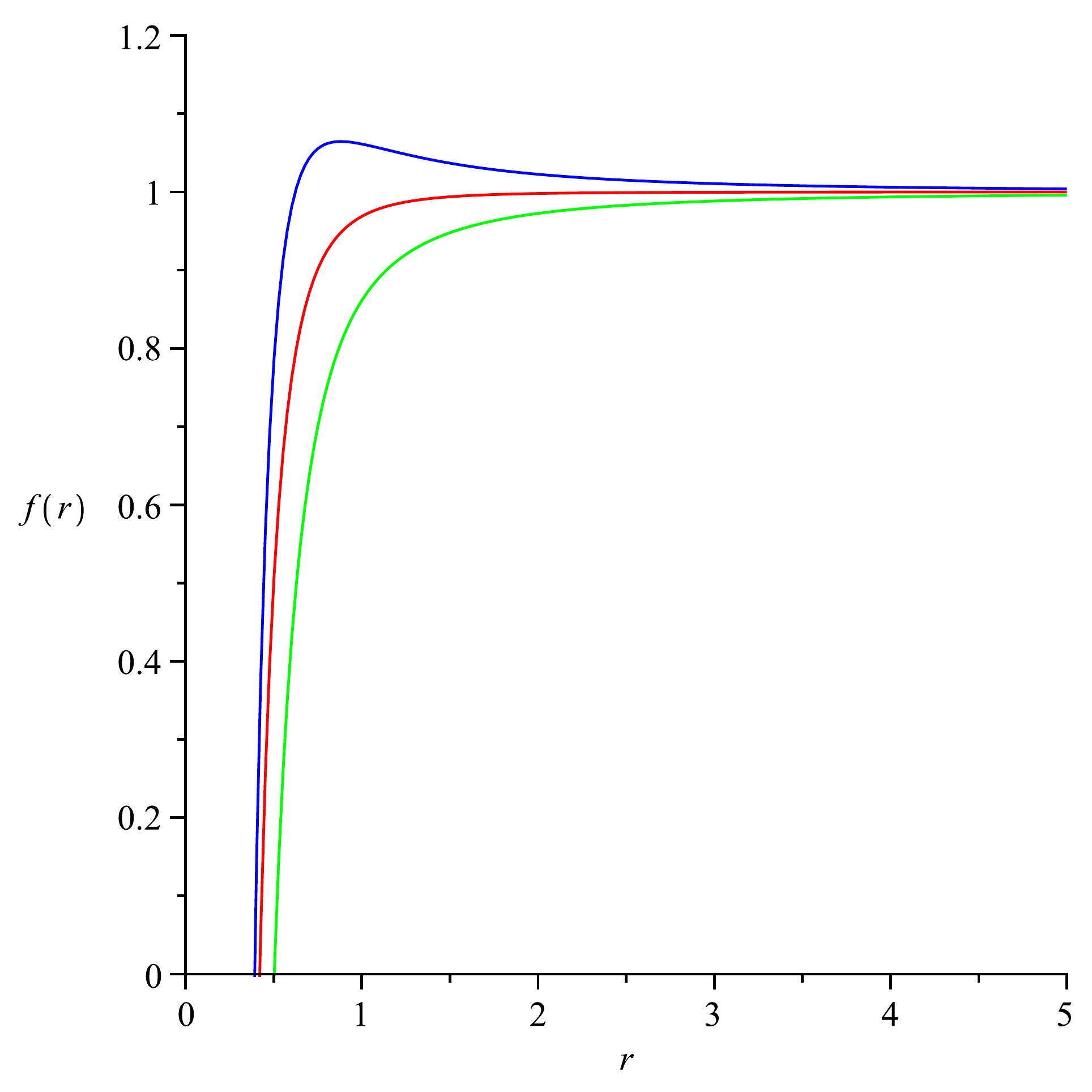}\qquad} {%
\includegraphics[width=.4\textwidth]{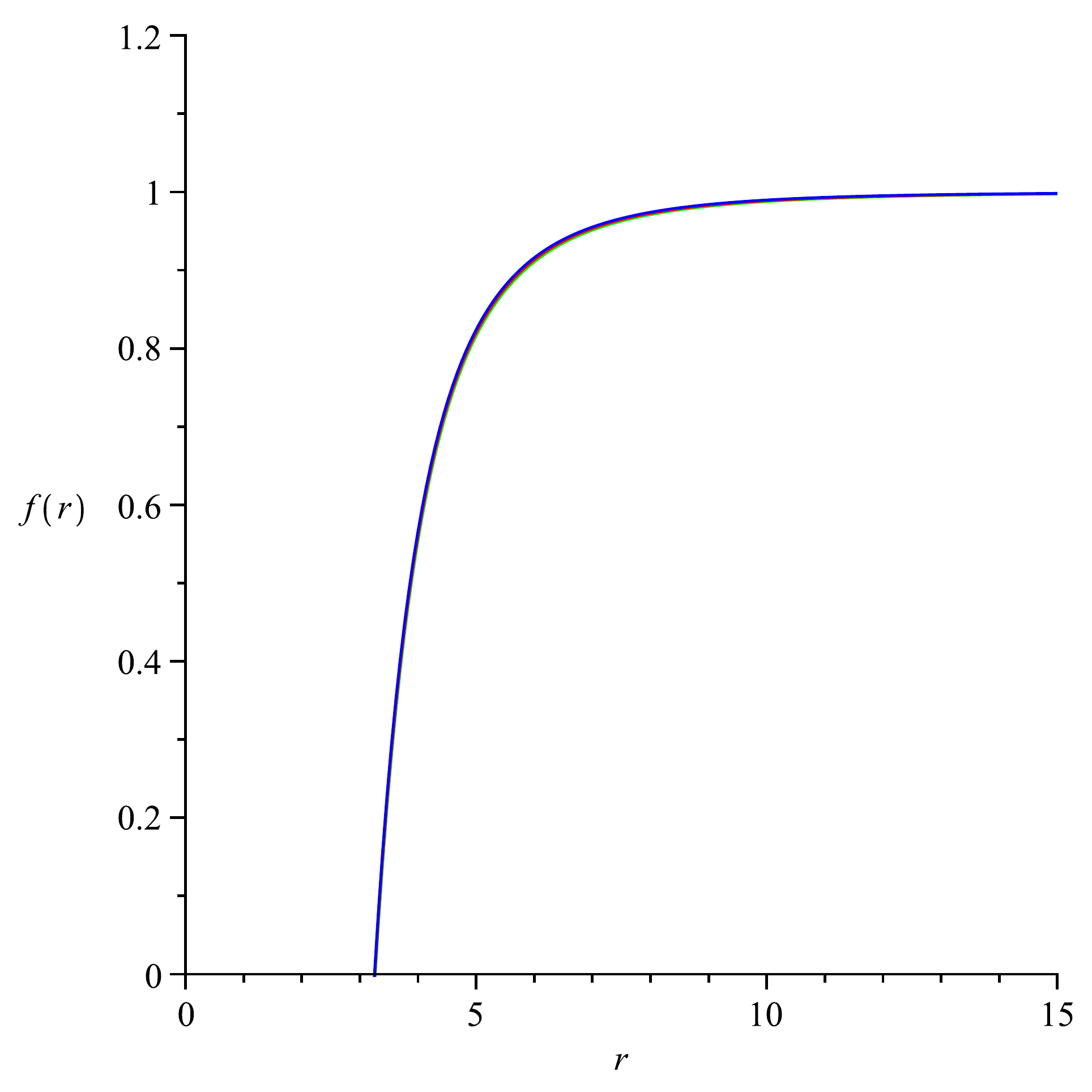}}
\caption{Metric functions $f(r)$ in 4-dimensions with: Left)
$Q=0.5$ for $k=-1$ (green), $k=0$ (red) and $k=1$ (blue). Right)
$Q=30$ for all $k$'s.} \label{fran4}
\end{figure}

For $n=2$ the metric is the same for all values of $k$ and the
solution is
\begin{equation}
h=1, \quad f^2=j^2=\frac{1}{g^2}=1-\frac{Q^2}{2r^2}, \quad
\kappa=\frac{Q}{r}\left(1-\frac{r_0}{r} \right)
\end{equation}
while in 4 dimensions ($n=3$) they differ for different $k$'s
\cite{Pang}:
\begin{equation}
h=1+\frac{k\ell^{2}}{20r^2}, \quad
f^2=j^2=\frac{1}{g^2}=1+\frac{k\ell^2}{10r^2}-\frac{3\ell^4k^2}{400r^4}-\frac{Q^2}{8r^4},
\quad \kappa=\frac{Q}{2r^{2}}\left(1-\frac{r^2_0}{r^2} \right)
\label{an4}
\end{equation}
One may note that in contrast to RN black holes, the
four-dimensional metric functions (\ref{an4}) have just one
horizon at:
\begin{equation}
r^{2}_{0}=-\frac{k\ell^2}{20}+\frac{\sqrt{4k^2\ell^4+50Q^2}}{20}.
\end{equation}
For small charges -- which means small black holes -- the
distinction between solutions with different $k$ but the same
charge are manifest. However  as charge increases the black hole
radius becomes larger and the distinction between metrics of
different $k$ with the same charge becomes very small, as shown in
Fig. (\ref{fran4}).

The other alternative occurs for $0<z<1$. In this case the first
two terms in eq. (\ref{EA}) would be canceled if one chooses:
\begin{equation}
Q^{2}=2z(1-z)\tilde{h}^{2}.\label{Qzless1}
\end{equation}
Now solving the remaining terms in (\ref{EA}) for $\tilde{h}$
yields:
\begin{equation}
\tilde{h}=\frac{(n-1)(n-2)(2-z)k\ell^{2}r^{n-3}}{z(2+z-2z^2)-z(2-z)n},\label{zless1}
\end{equation}
which is a constant for charged solutions if $n=3$. Substituting
Eqs. (\ref{Qzless1}) and (\ref{zless1}) and solving the field
equations for $n=3$, the solutions are given by:
\begin{eqnarray}
&&h=1-\frac{(2-z)k\ell^{2}}{z(z^2-2z+2)r^{2}}, \quad
f^2=j^2=\frac{1}{g^2}=1+\frac{k\ell^{2}}{(z^2-2z+2)r^2}, \nonumber \\
&&\kappa=\pm\sqrt{\frac{2(1-z)k^2\ell^4}{z(z^2-2z+2)^2}}\frac{1}{r^{2}}\left(1-\frac{r^{z-2}_0}{r^{z-2}}\right)
\label{an4b}
\end{eqnarray}
Therefore the 4 dimensional metric function with $0<z<1$ is a unit
function for $k=0$, a naked singularity for $k=1$ and a black hole
with one horizon located at $r_{0}^{2}=\ell^2/(z^2-2z+2)$ for
$k=-1$.

\section{Solutions at large $r$}

\label{larger}

In the previous section we wrote down the field equations to
higher dimensions and arbitrary $z$ (equations
(\ref{Eqf})--(\ref{Eqh})), and here we review the general form of
the asymptotic behaviour of the solutions  \cite{Zingg2}.

We begin with linearizing the system in ($n+1$)-dimensions. Since
we require the general metric (\ref{metric}) to asymptotically
approach  the Lifshitz one (\ref{asmet}), we investigate the
behaviour at large $r$ by applying a small perturbation for the
fields
\begin{eqnarray}
&&f(r)=1+w f_1(r),  \nonumber \\
&&g(r)=1+w g_1(r),  \nonumber \\
&&j(r)=1+w j_1(r),  \nonumber \\
&&h(r)=1+w h_1(r).  \label{infexh}
\end{eqnarray}

In the charged case we must also consider constraints on the
behaviour of the gauge field to ensure Lifshitz asymptotics. We
first note that the quantity $Q$ is proportional to the electric
charge per unit volume, since
\begin{equation}
\mathcal{Q}=\frac{1}{16\pi \Omega _{k}}\int_{S}\ {}^{\ast
}F=\frac{1}{16\pi
\Omega _{k}}\int d\Omega _{k}r^{n-1}n^{\mu }F_{\mu \nu }u^{\nu }=\frac{1}{%
16\pi \Omega _{k}}\int d\Omega _{k}r^{n-1}\frac{Qr^{1-z}}{r^{n-z}}=\frac{Q}{%
16\pi }  \label{Qconserve}
\end{equation}
where $u^{\mu }$ and $n^{\mu }$ are the unit timelike and
spacelike normals to a sphere of radius $r$.

Since $f(r)$ does not contribute to the equations for $g(r)$, $h(r)$ and $%
j(r)$ we can first study the set of equations involving
$\{g,h,j\}$. Inserting the perturbative expansion (\ref{infexh})
into equations (\ref{Eqg}-\ref{Eqh}), we obtain the equations for
small perturbations
\begin{eqnarray}
r\frac{d}{dr}\pmatrix{\delta g\cr\delta h\cr\delta j} &=&
\pmatrix{ -n && z(z-1)/(n-1) && z-1 \cr n-1 && 1-n && n-1 \cr
-(n+z-2) && z(n+z-2)/(n-1) && 1-2z }\pmatrix{ g_1 \cr h_1 \cr
j_1} \nonumber\\
&&\qquad  +\left( \frac{Q^2}{4(n-1)r^{2n-2}} +\frac{%
(n-2)k}{2r^{2}}\right) \pmatrix{1\cr0\cr1} \label{Eqp}
\end{eqnarray}
where we have included the Maxwell gauge field as a first
order perturbation which means that we substitute
$Q^2/r^{2(n-1)}=wQ^2/r^{2(n-1)}$, since its falloff may be slower
than other terms in the metric functions. We have also  rescaled
$r\rightarrow r/\ell$. Note that for $k\neq 0$ a universal
$1/r^{2}$ mode also contributes to the metric functions.

A detailed discussion of the large $r$ expansion is given in the
appendix. The eigenvalues of the matrix of coefficients is
obtained via straightforward calculation
\begin{equation}
\pmatrix
{z+n-1\cr\cr\frac{z+n-1}{2}+\frac{\sqrt{9z^{2}-2(3n+1)z+(n^{2}+6n-7)}}{2}\cr%
\cr\frac{z+n-1}{2}-\frac{\sqrt{9z^{2}-2(3n+1)z+(n^{2}+6n-7)}}{2}}
\label{eigen}
\end{equation}
indicating there are three independent eigenmodes.

The two upper modes in (\ref{eigen}) are always decaying. However
the lowest mode, depending on the value of $z$, will either be a
growing mode if $z>n-1$, a zero mode (independent of $r$) if
$z=n-1$, or a decaying mode if $z<n-1$. Then if one looks for the
solutions that asymptotically approach a Lifshitz fixed point, one
has to discard the lowest mode in (\ref{eigen}) for $z\geq n-1$ by
demanding the amplitude of this mode vanishes at large $r$.

Removing zero or growing modes can be done numerically in both
uncharged or charged solutions by fine-tuning initial values upon
solving the field equations. It is obvious that one has to remove
the growing mode for $z>n-1$. The zero mode for $z=n-1$ becomes a
marginally growing mode when non-linear corrections are included
\cite{Daniel}. Such a mode with positive amplitude cannot approach
the Lifshitz geometry (\ref{asmet}) asymptotically and therefore
must be removed. On the other hand, a zero mode with negative
amplitude will slowly decay; in this case a consideration of the
evolution
of $f(r)$, (ignored thus far), indicates that it goes to zero as $%
r\rightarrow \infty$ again yielding unacceptable asymptotic
behaviour.

For $z<n-1$ all three eigenmodes decay, and so no fine-tuning of
initial values is required  if one simply wishes to obtain
solutions that asymptote to the Lifshitz metric (\ref{asmet}). For
a given event horizon size we have a family of solutions that are
all asymptotic to the Lifshitz metric (\ref {asmet}), but with
different fall-off rates. By an appropriate choice of boundary
terms, such solutions will have finite energy
\cite{Saremi,Copsey}.
\begin{figure}[tbp]
\centering
{\includegraphics[width=.4\textwidth]{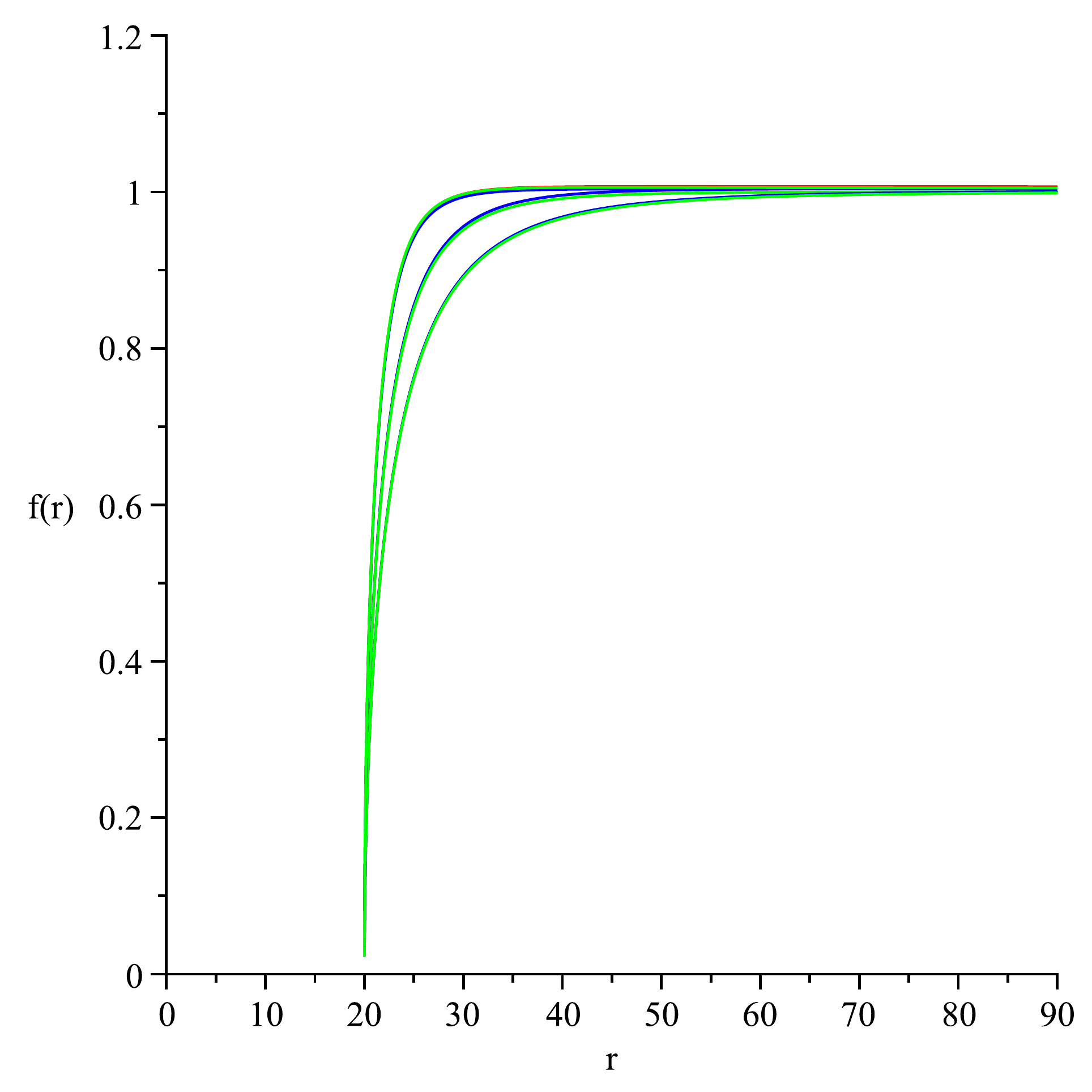}\qquad} {%
\includegraphics[width=.4\textwidth]{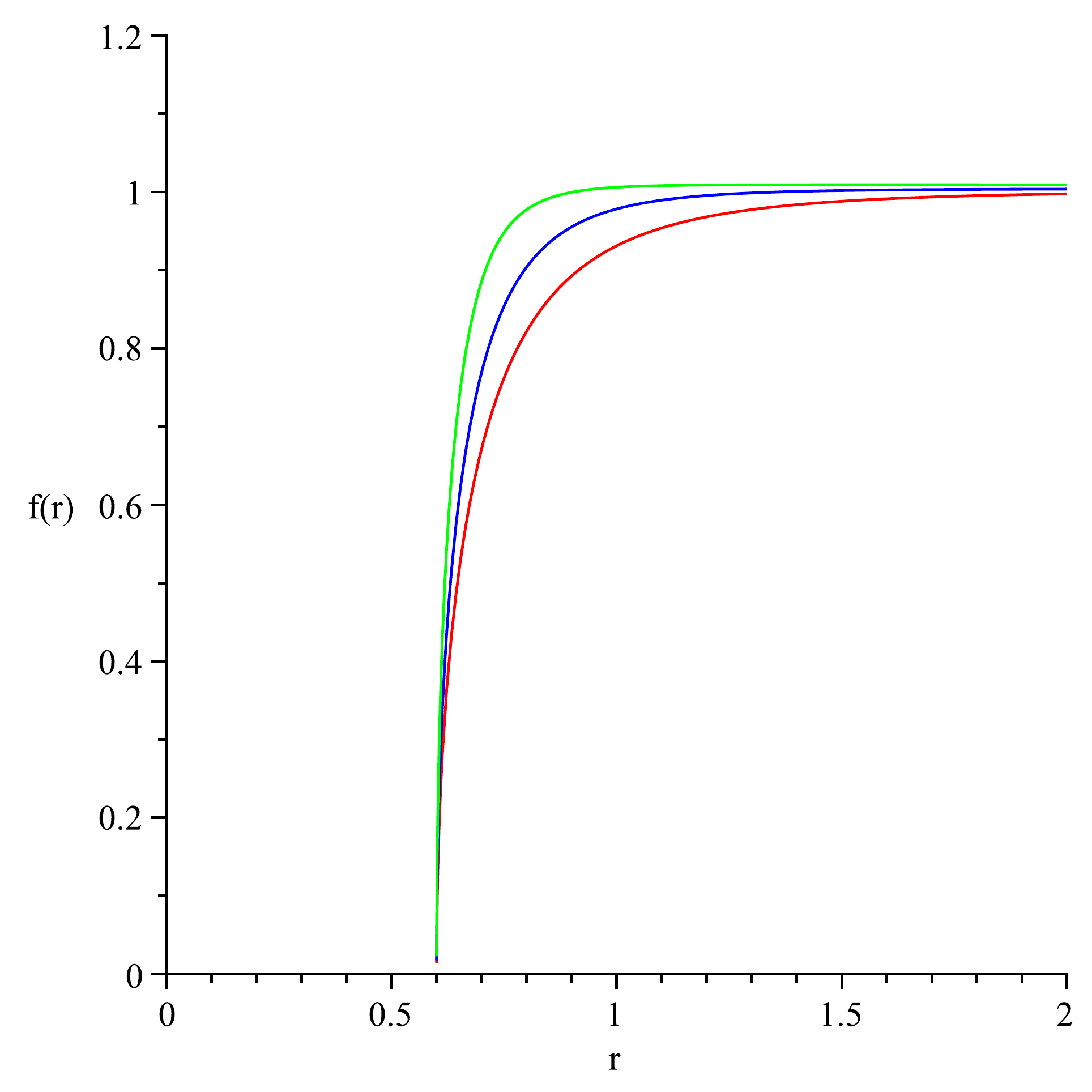}}
\caption{Metric functions $f(r)$ for zero modes in 4, 5 and
7-dimensions from bottom to top
respectively. Left) $r_{0}=20$ for all $k$. Right) $r_{0}=0.6$ for $%
k=0$.} \label{f1r20}
\end{figure}

\section{Near horizon expansion}

In order to investigate the near horizon behavior of the
solutions, consider the following expansions for the fields:
\begin{eqnarray}
&&f(r)=f_{0}\sqrt{r-r_{0}}(1+f_{1}(r-r_{0})+f_{2}(r-r_{0})^{2}+\cdots
),
\nonumber \\
&&g(r)=\frac{g_{0}}{\sqrt{r-r_{0}}}(1+g_{1}(r-r_{0})+g_{2}(r-r_{0})^{2}+%
\cdots ),  \nonumber \\
&&j(r)=j_{0}\sqrt{r-r_{0}}(1+j_{1}(r-r_{0})+j_{2}(r-r_{0})^{2}+\cdots
),
\nonumber \\
&&h(r)=h_{0}(1+h_{1}(r-r_{0})+h_{2}(r-r_{0})^{2}+\cdots ),
\label{Eqnh}
\end{eqnarray}
where $r_{0}$ is the horizon radius. All other constants in the
series solutions (\ref{Eqnh}) can be obtained in terms of $r_{0}$
and $h_{0}$ as well but we omit them here because they are too
lengthy.

Inserting this into eqs. (\ref{Eqf}-\ref {Eqh}) and demanding the
coefficients for each power of $r-r_{0}$ vanish determines all
constants in terms of $r_{0}$ and $h_{0}$. For example $g_{0}$ is
given by
\begin{equation}
g_{0}=r_{0}^{1/2}\sqrt{\frac{2(n-1)}{%
r_{0}^{2-2n}(Q_{c}^{2}-Q^{2})-2z(z-1)h_{0}^{2}}} \label{g0}
\end{equation}
where
\begin{equation}
Q_{c}^{2}\equiv \frac{2}{r_{0}^{4-2n}}\left\{ \left[
(z-1)^{2}+(z-2)n+n^{2}\right] r_{0}^{2}+(n-1)(n-2)k\ell
^{2}\right\} . \label{Qc}
\end{equation}
\begin{figure}[tbp]
\centering
{\includegraphics[width=.4\textwidth]{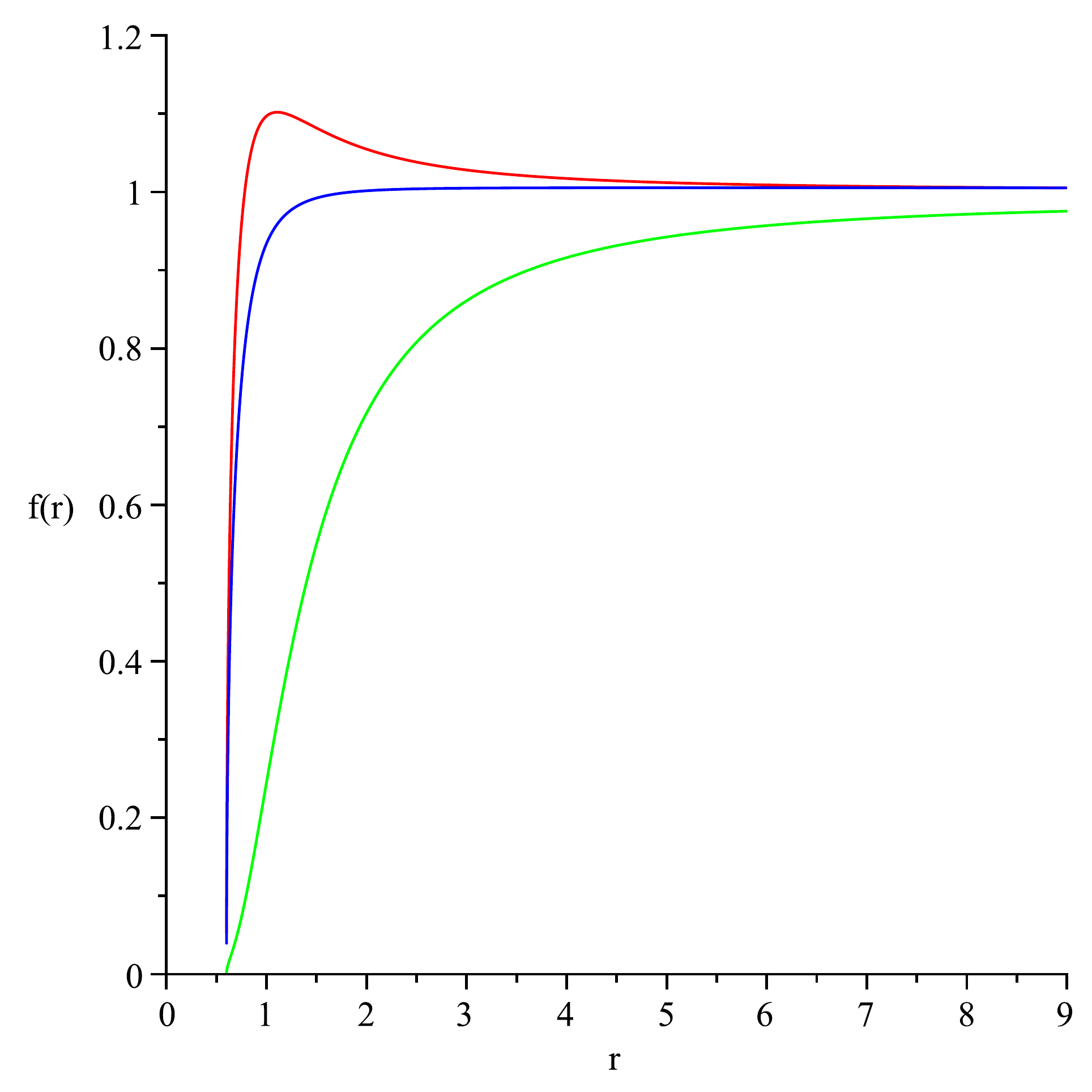}\qquad} {%
\includegraphics[width=.4\textwidth]{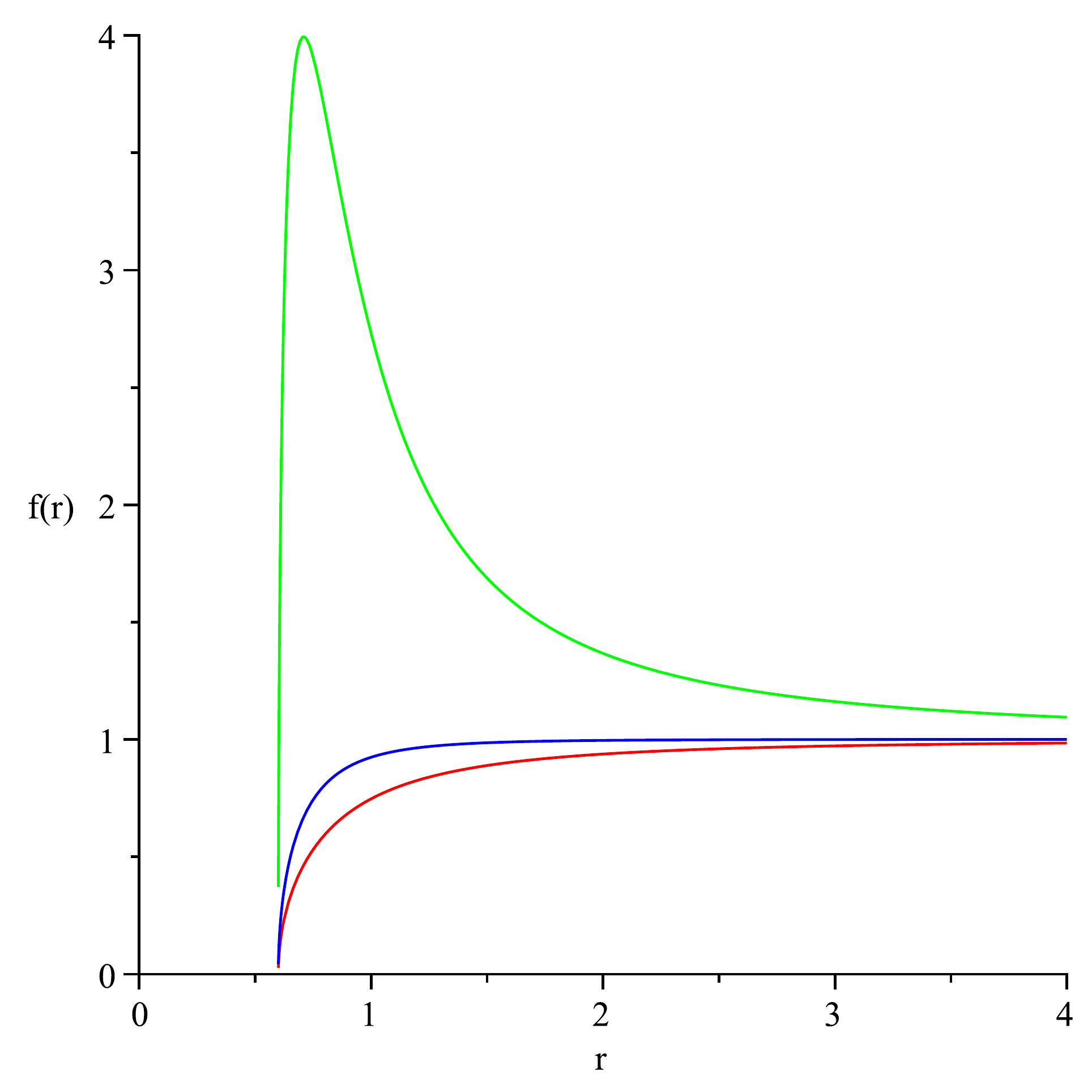}}
\caption{Metric functions $f(r)$ for zero modes with $r_{0}=0.6$
in 4 (red), 5 (blue) and 7 (green)-dimensions for: Left) $k=1$;
Right) $k=-1$.} \label{kpmrs}
\end{figure}
Requiring $g_{0}$ to be real we obtain the constraint
\begin{equation}
|Q|\leq \sqrt{Q_{c}^{2}-2z(z-1)h_{0}^{2}{r_{0}^{2n}}}
\label{Qcons}
\end{equation}
generalizing a similar constraint obtained for $Q=0$ black holes
\cite {Daniel,Robb}. Indeed, for $Q=0$ the constraint
(\ref{Qcons}) reduces to
\begin{equation}
\left[ (z-1)^{2}+(z-2)n+n^{2}-z(z-1)h_{0}^{2}\right] r_{0}^{2}+(n-1)(n-2)k%
\ell ^{2}\geq 0  \label{h0cons}
\end{equation}
imposing either an upper bound on $|h_{0}|$ if $k=0,1$ or imposing
a lower bound of
$r_{0}>\sqrt{\frac{(n-1)(n-2)}{(z-1)^{2}+(z-2)n+n^{2}}}$ if
$k=-1$.


For $Q\neq 0$ equation (\ref{Qcons}) is more usefully written as
\begin{equation}
\frac{(z-1)^{2}+(z-2)n+n^{2}-z(z-1)h_{0}^{2}}{(n-1)(n-2)}r_{0}^{2}+k\ell
^{2}\geq \frac{Q^{2}}{2\ell ^{2}(n-1)(n-2)r_{0}^{2(n-1)}}
\label{Qcons2}
\end{equation}
or alternatively as $\alpha X+k>X^{1-n}$, where
\begin{equation}
\alpha
=\frac{(z-1)^{2}+(z-2)n+n^{2}-z(z-1)h_{0}^{2}}{(n-1)(n-2)}\left(
\frac{Q^{2}\ell ^{-2(n+1)}}{2(n-1)(n-2)}\right) ^{\frac{1}{n-1}}
\label{Qcons3}
\end{equation}
upon setting $r_{0}^{2}=X\ell ^{2}\left( \frac{Q^{2}\ell ^{-2(n+1)}}{%
2(n-1)(n-2)}\right) ^{\frac{1}{n-1}}$. We see that analysis of
this equation involves finding the intersection points of a
straight line in $X$ with a curve behaving as $X^{1-n}$. The line
intersects the origin for $k=0$, and has an intercept at $\pm 1$
for $k=\pm 1$. For $h_{0}\rightarrow 0$ the line always intersects
the $X^{1-n}$ curve, and so any values of $X$ larger than this are
admissible. Hence for any given $Q$ there is a lower bound on the
size of the black hole regardless of the value of $k$. The lower
bound is largest for $k=-1$ and smallest for $k=1$. As $h_{0}$
increases the slope of the line decreases, and so this lower bound
increases. For $k=0,-1$ it
becomes infinite as $\alpha \rightarrow 0$, i.e. $h_{0}\rightarrow \frac{%
(z-1)^{2}+(z-2)n+n^{2}}{z(z-1)}$. Indeed for small $\alpha $ we
find
\[
X\geq \frac{1}{\alpha }\left( 1-(n-1)\alpha ^{n-1}-\frac{(n-2)(n-1)}{2}%
\alpha ^{2n-2}+\frac{(4n^{2}-8n+3)(n-1)}{3}\alpha ^{3n-3}+\cdots
\right)
\]
and we see that the lower bound on $X$ diverges as $\alpha
\rightarrow 0$.

However for $k=1$ the situation is quite different. As
$h_{0}\rightarrow \frac{(z-1)^{2}+(z-2)n+n^{2}}{z(z-1)}$ the lower
bound on $X$ remains finite. As $h_{0}$ becomes even larger , a
larger lower bound on $X$ appears since now $\alpha <0$. As
$h_{0}$ increases, $\alpha $ becomes more negative, and the lower
bound on $X$ continues to decrease. Eventually a limit of
$X>n^{\frac{1}{n-1}}$ is reached at which the line is tangent to
the $X^{1-n}$ curve, where $\alpha =(1-n)n^{\frac{n}{1-n}}$, or
\[
\frac{z(z-1)h_{0}^{2}-(z-1)^{2}+(z-2)n+n^{2}}{(n-1)^{2}(n-2)}=\left( \frac{%
Q^{2}n^{n}\ell ^{-2(n+1)}}{2(n-1)(n-2)}\right) ^{\frac{1}{1-n}}
\]
For values of $h_{0}$ larger than this, no black hole solutions
exist; instead there is a naked singularity.

Alternatively if we fix $r_{0}$ then demanding $g_{0}$ to be real implies $%
Q_{c}^{2}>0$ (and so $|Q|<|Q_{c}|$) as well as
\begin{equation}
\mid h_{0}\mid <\sqrt{\frac{\ell ^{2}r_{0}^{2-2n}(Q_{c}^{2}-Q^{2})}{%
2z(z-1)L^{4}}}.
\end{equation}
Positivity of $Q_{c}^{2}$ is always satisfied if $k=0,1$. However
for $k=-1$ it is satisfied provided $r_{0}$ respects a lower bound
as
\begin{equation}
r_{0}>\sqrt{\frac{(n-1)(n-2)}{n^{2}+n(z-2)+(z-1)^{2}}}.
\end{equation}
The value $Q\to Q_c$ is the extremal limit of the charged Lifshitz
black hole \cite{Zingg,Zingg2}, defined in eq. (\ref{Qc}); it
corresponds to the value of $Q$ for which the temperature
vanishes.

\section{Numeric Solutions in ($n+1$)-dimensions}

In order to find numeric solutions for the system of ODE's
(\ref{Eqf}-\ref {Eqh}) we apply the shooting method by adjusting
some initial values for the fields $f(r)$, $g(r)$, $j(r)$ and
$h(r)$. To find the initial values we
employ the near horizon expansions (\ref{Eqnh}) by substituting $%
r_{0}+\varepsilon$ for $r$ where $\varepsilon\ll1$, choosing some
values for $r_{0}$ and $h_{0}$ such that after applying numerical
method the metric and field functions approach unity at large $r$.

\subsection{Uncharged Einstein-Lifshitz Solutions}

In order to study uncharged Einstein-Lifshitz solutions we put
$Q=0$ in ODE's (\ref{Eqf}-\ref{Eqh}) and use the numerical method
to find the solutions. We already explained there are different
situations depending on the value of critical exponent $z$:

\textbf{I)} We begin with an examination of situations where the
zero mode is present, which implies $z=n-1$. For a given horizon radius $%
r_{0}$ the solution is unique, requiring a fine-tuning of initial
values to approach the Lifshitz metric (\ref{asmet}) at large $r$.
We find that the solutions are similar to the $n=3$ case
\cite{Robb}: for each $n$, when $k=0$
the same value of $h_{0}$ yields black hole solutions for all values of $%
r_{0}$. Specifically, $h_{0}=1.3737, 1.3344, 1.3084$ in 4, 5 and 7
dimensions respectively. For $k=1 $ ($-1$) it is necessary to
systematically
adjust the value of $h_{0}$ upward (downward) relative to the value for $k=0$%
. Also, for each $n$ large black holes are almost
indistinguishable for different choices of $k$, see Fig.
(\ref{f1r20}-left). Distinctions start to appear between solutions
of the same dimension but different $k$ as the horizon radius
$r_{0}$ gets smaller.

\begin{figure}[tbp]
\centering {\includegraphics[width=.4\textwidth]{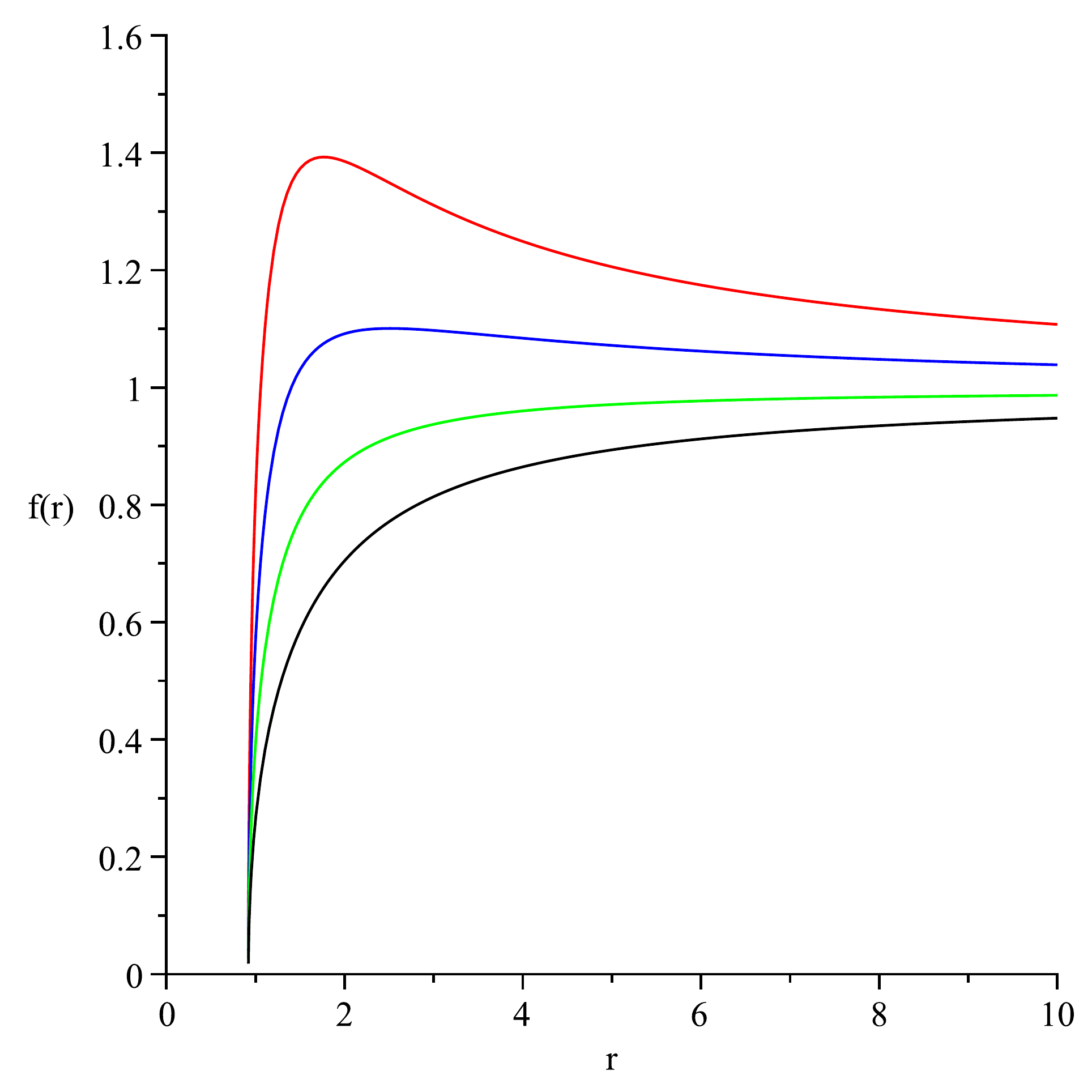}}
\centering {\includegraphics[width=.4\textwidth]{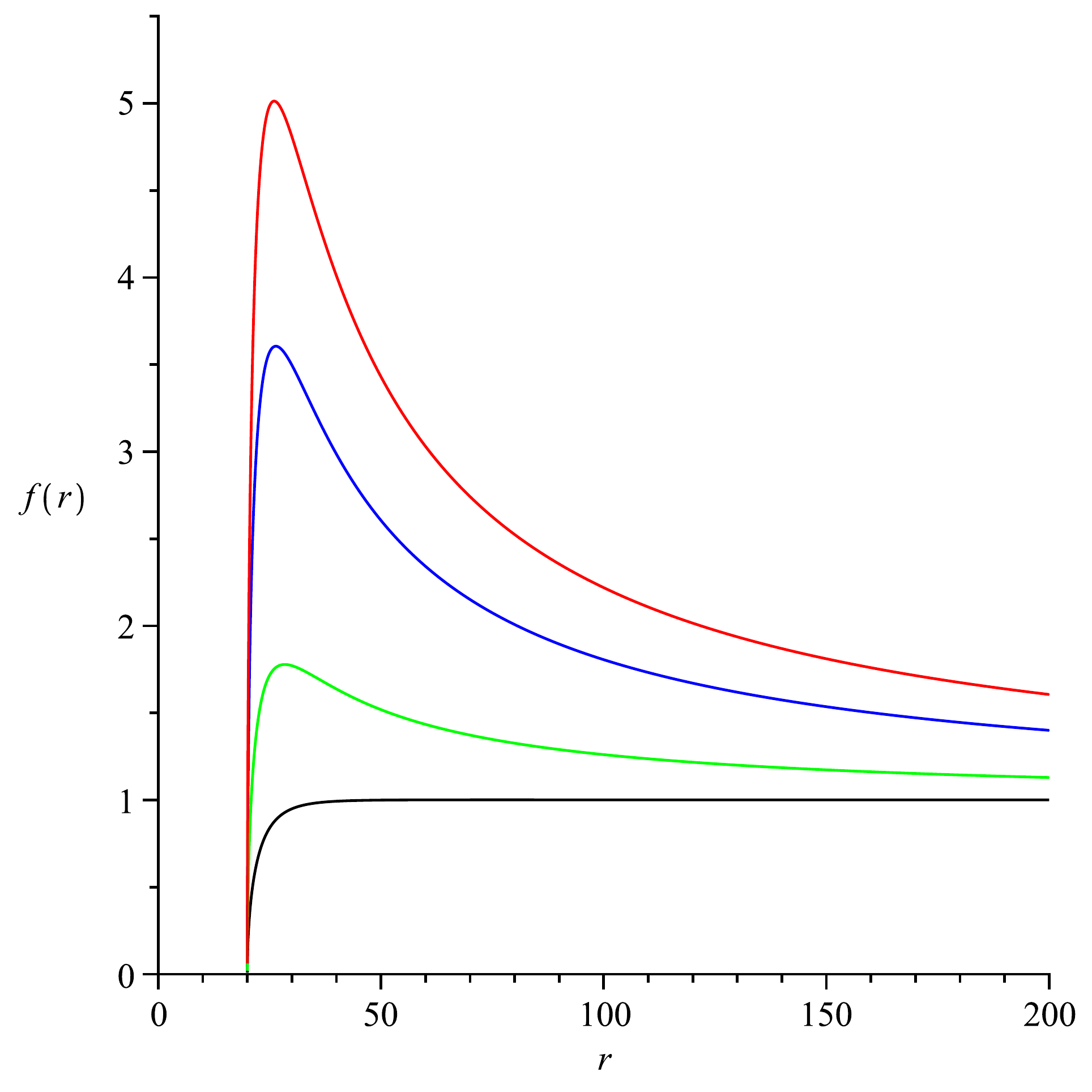}}
\caption{Metric functions $f(r)$ for $k=-1$ in 7-dimensions with
$z=2$: {Left)} $r_{0}=0.92$.  {Right)} $r_{0}=20$
while the massive gauge field strength $h_{0}$ increases and
temperature decreases from top to bottom in both cases. }
\label{k-1n7}
\end{figure}

We can also probe the behaviour of our solutions as a function of
dimensionality. Looking at the left-hand side of Fig.
(\ref{f1r20}), we see that for large $r_{0}$, where the solutions
are almost the same different
values of $k$, as the dimension increases the solutions approach (\ref{asmet}%
) more rapidly as $r$ increases. However the overall behaviour of
the solutions is essentially the same: the metric function $f(r)$
starts from zero at the horizon radius and monotonically increases
to asymptotically approach unity at large $r$. The situation is
the same for small black holes when $k=0$, shown at the right in
Fig. (\ref{f1r20}).

However, for small black holes with $k=1$ or $k=-1$ the situation
differs, as depicted in Fig. (\ref{kpmrs}). For $k=1$ in
4-dimensions, the metric function $f(r)$ starts from zero at the
horizon radius increasing to reach a maximum above unity, then
decreasing to approach unity at large $r$. However in dimensions
greater than 4, the metric function $f(r)$ starts from zero at the
horizon radius and monotonically increases to asymptotically
approach unity as $r$ goes to infinity. For $k=-1$ the situation
is completely vice-versa. Examples are shown in Fig.
(\ref{kpmrs}), with $k=1$ on the left and $k=-1$ on the right.
\begin{figure}[tbp]
\centering
{\includegraphics[width=.4\textwidth]{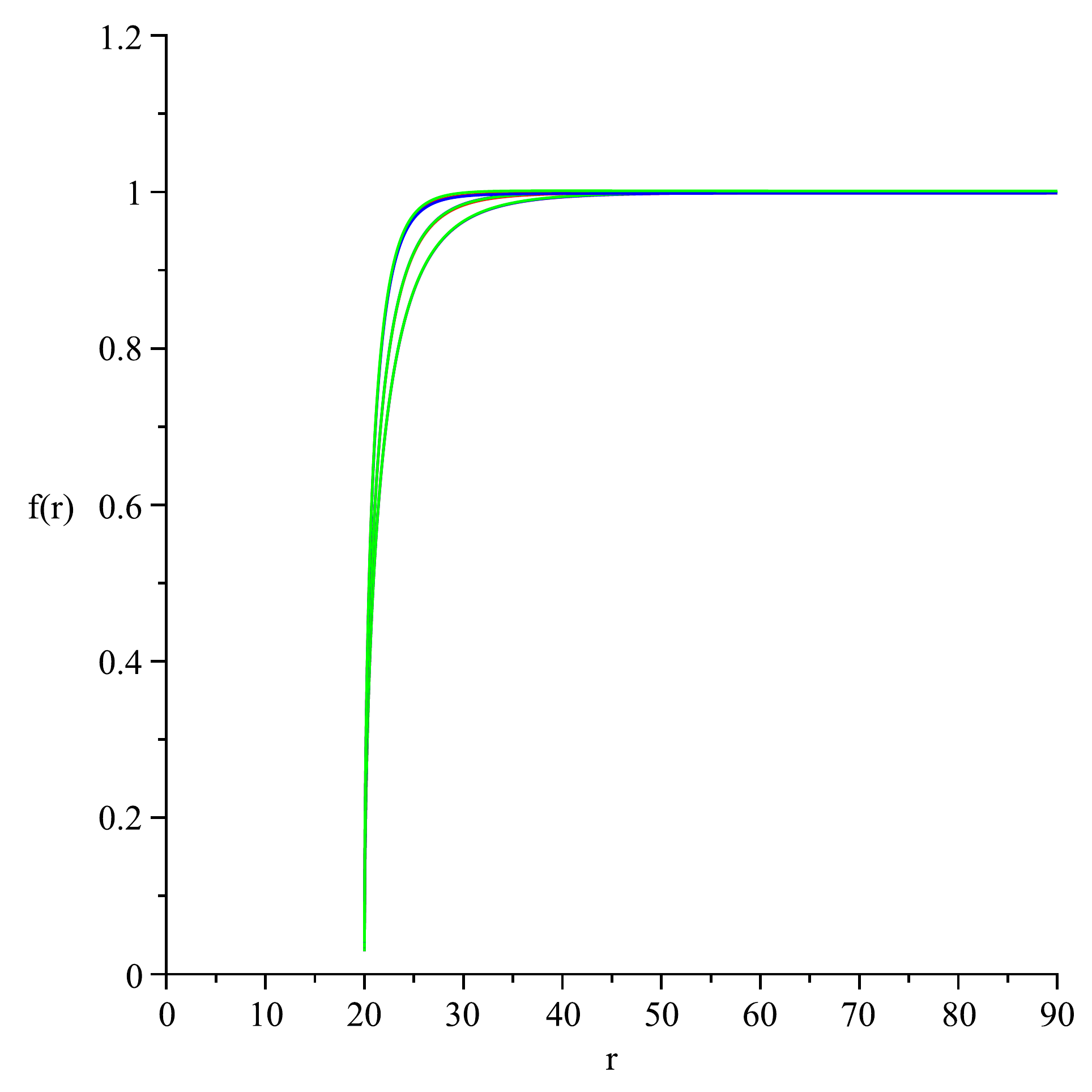}\qquad} {%
\includegraphics[width=.4\textwidth]{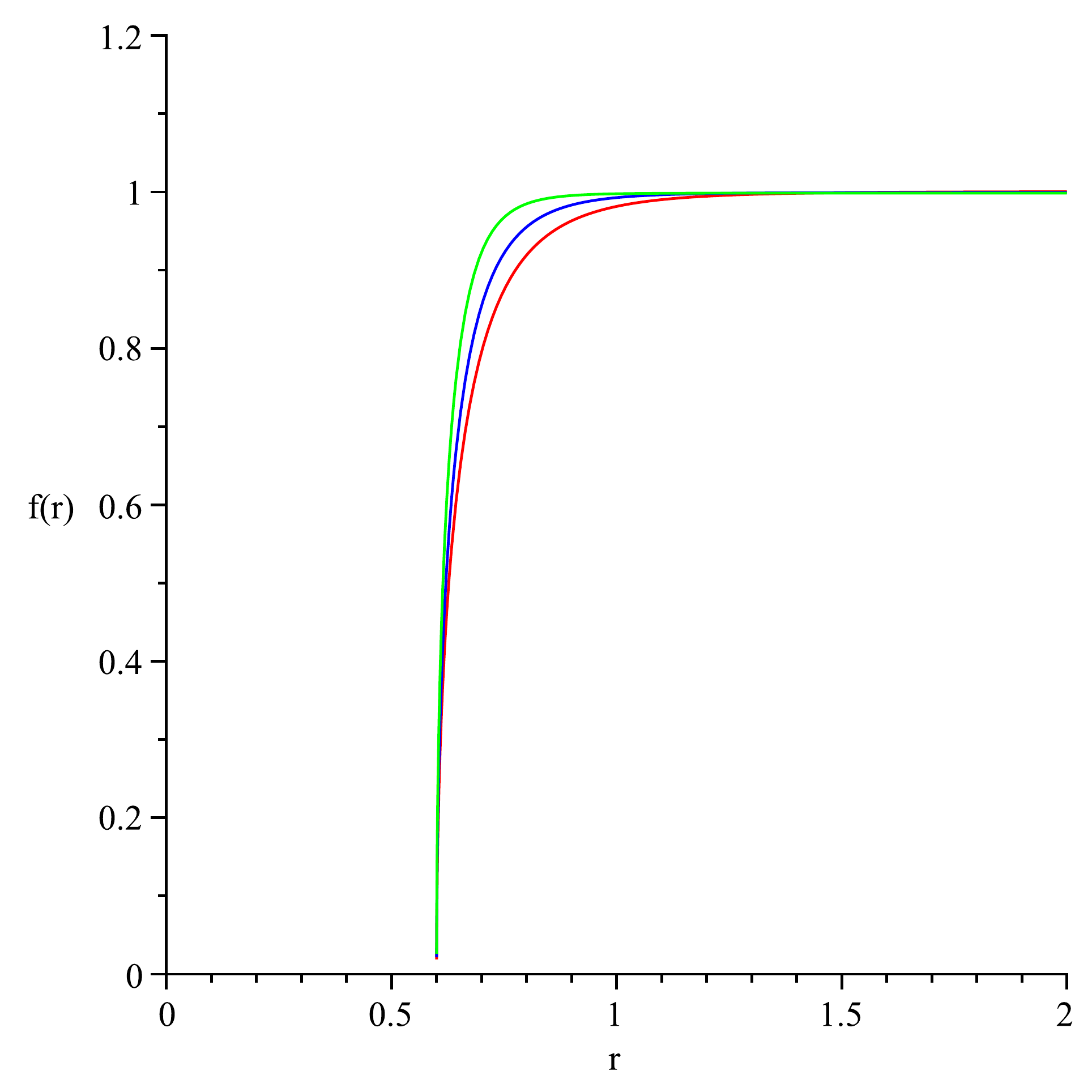}}
\caption{Metric functions $f(r)$ in 4, 5 and 7-dimensions with
$z=$4, 5 and 7 from bottom to top, respectively. Left)
$r_{0}=20$ and all values of $k$'s.  For a given dimension the different values of $k$ lie
almost exactly on the same curve.
 Right) $r_{0}=0.6$ and $k=0$ .}
\label{f2r20}
\end{figure}

\textbf{II)} $z<n-1$:   For the series solutions at large $r$
all three eigenmodes resultant from the small perturbation are
decaying, and so no fine-tuning is required. In this case it is
possible to find solutions that all approach unity at large $r$ by
fixing $r_{0}$ and then searching for a value of $h_{0}$ such that
the boundary conditions are satisfied. We find that $h_{0}$ is not
unique; in other words there is a continuous spectrum of values of
$h_{0}$ that all respect the boundary conditions. Numerical
exploration shows that for strong massive gauge field, i.e. a
larger value of $h_{0}$, the metric function $f(r)$ starts from
zero at the horizon radius and monotonically increases to
asymptotically reach unity. However, for a smaller value of
$h_{0}$, $f(r)$ starts from zero at the horizon radius and as $r$
increases, passes through a maximum value greater than unity, and
then approaches unity as $r$ goes to infinity. Moreover, black
holes with stronger massive gauge field are colder than those with
weaker ones. For example, in 7 dimensions with $z=2$ and $k=-1$
then, for a fixed $r_{0}$, there is a family of solutions that
have qualitatively different behaviour compared to each other but
all are asymptotic to the Lifshitz background at large $r$. An
illustration of this is in Fig. (\ref {k-1n7}) for both small and
large black holes. We find the same situation in other dimensions
with different choices of $z$ and $k$ provided $z<n-1$.
\begin{figure}[tbp]
\centering
{\includegraphics[width=.4\textwidth]{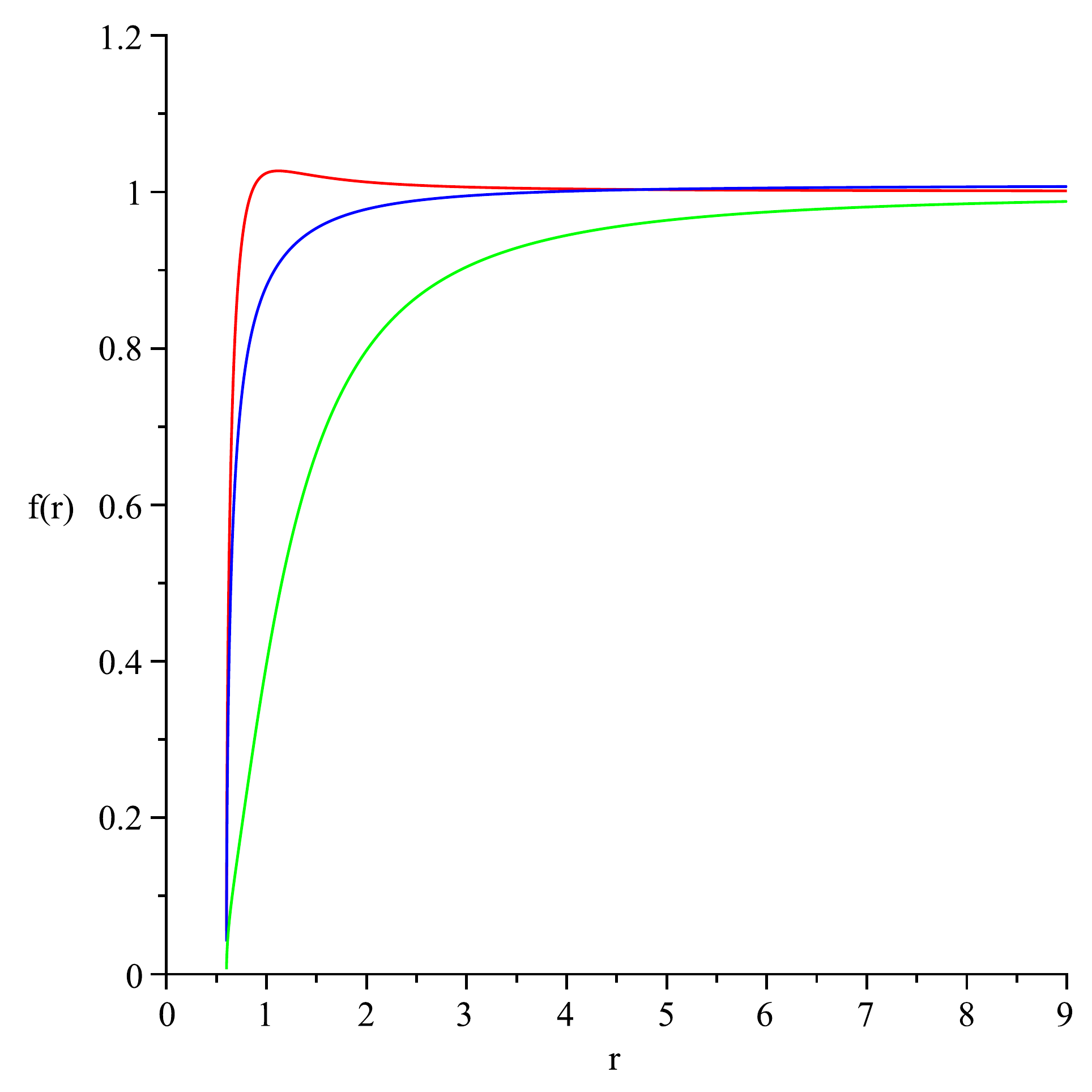}\qquad} {%
\includegraphics[width=.4\textwidth]{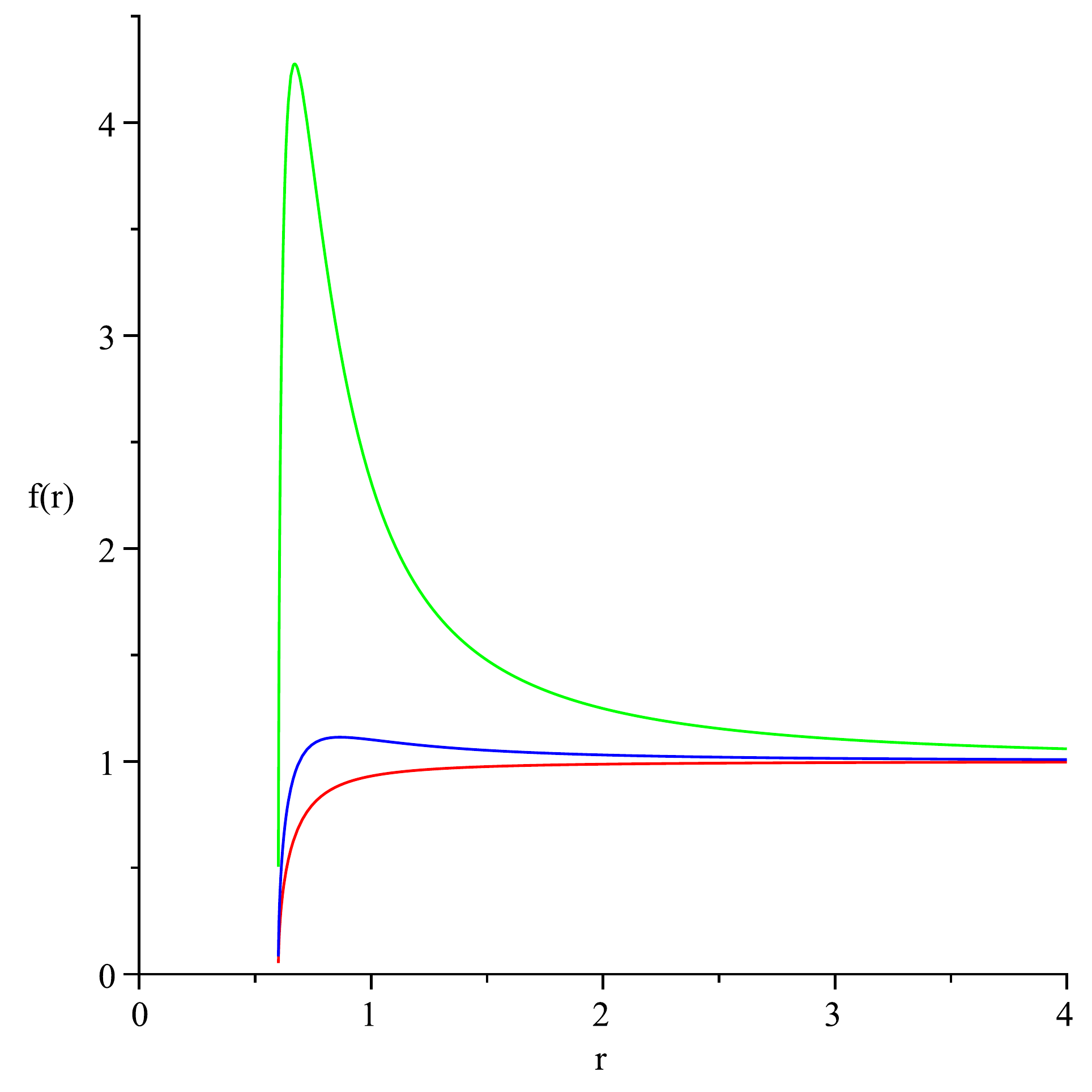}}
\caption{Metric functions $f(r)$ with $r_{0}=0.6$ in 4 (red), 5
(blue) and 7 (green)-dimensions and $z=4,5$ and $7$ respectively.
Left) $k=1$. Right) $k=-1$.} \label{kpmrsgmode}
\end{figure}

\textbf{III)} $z>n-1$: In this case one eigenmode amongst three is
growing. To have solutions asymptotic to (\ref{asmet}) the initial
values of the parameters must be adjusted such that the amplitude
of this mode is removed otherwise the fields diverge so fast at
large $r$. Compared to the zero mode the fine-tuning is
numerically more delicate -- approaching (\ref{asmet}) at large
$r$ needs to more accuracy in digits of $h_{0}$. However, the
results are similar to the zero mode case: for each $n$ large
black holes are almost indistinguishable for different choices of
$k$ and the metric function $f(r)$ reaches unity faster as the
dimension of spacetime increases, as shown in the left side of
Fig. (\ref{f2r20}). Distinctions start to appear between
solutions of the same dimension but different $k$ as the horizon radius $%
r_{0}$ gets smaller, illustrate in the right side of Figs.
(\ref{f2r20}) and in Fig. (\ref{kpmrsgmode}).

\subsection{Charged Einstein-Lifshitz Solutions}

\begin{figure}[tbp]
\centering {\includegraphics[width=.4\textwidth]{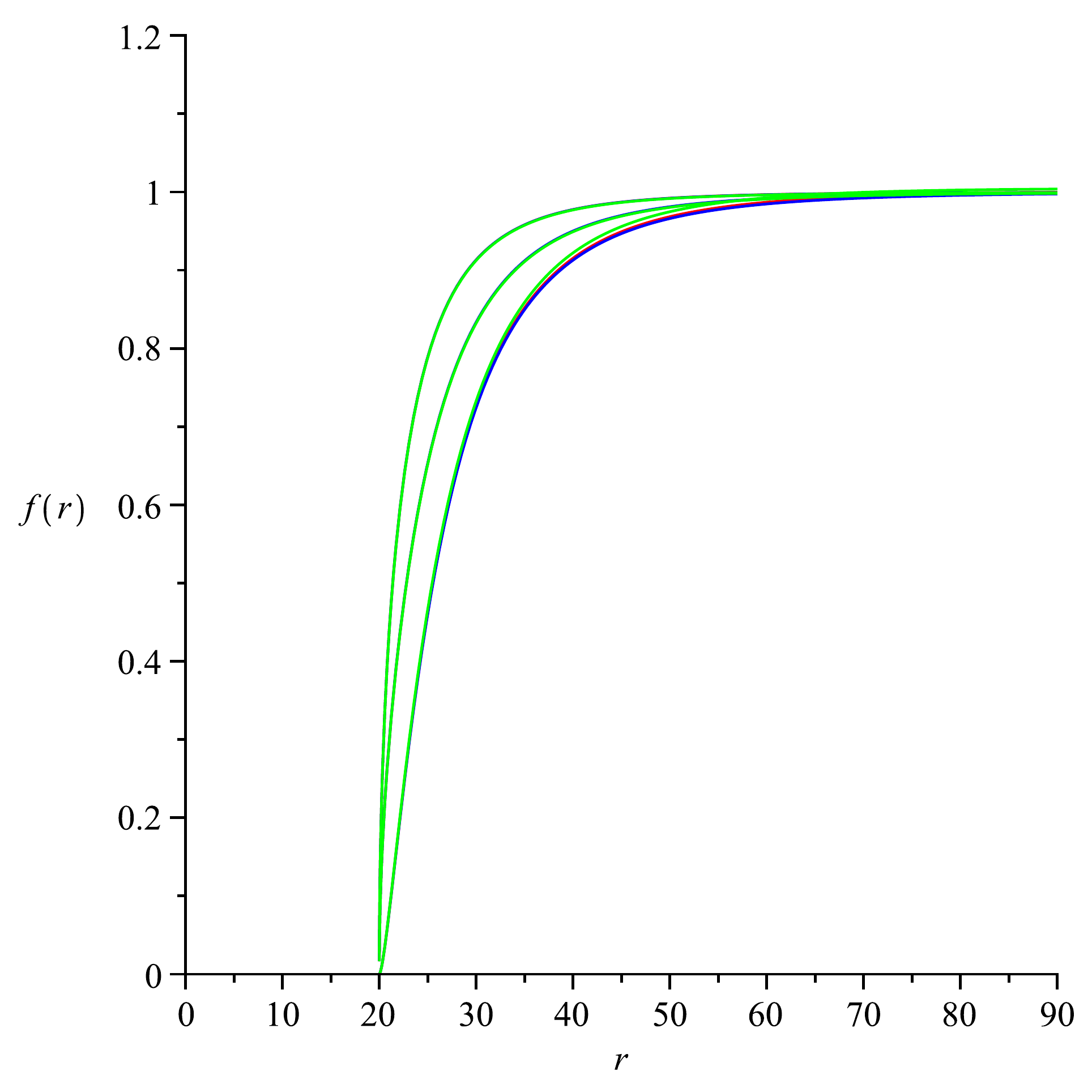}}\qquad{%
\includegraphics[width=.4\textwidth]{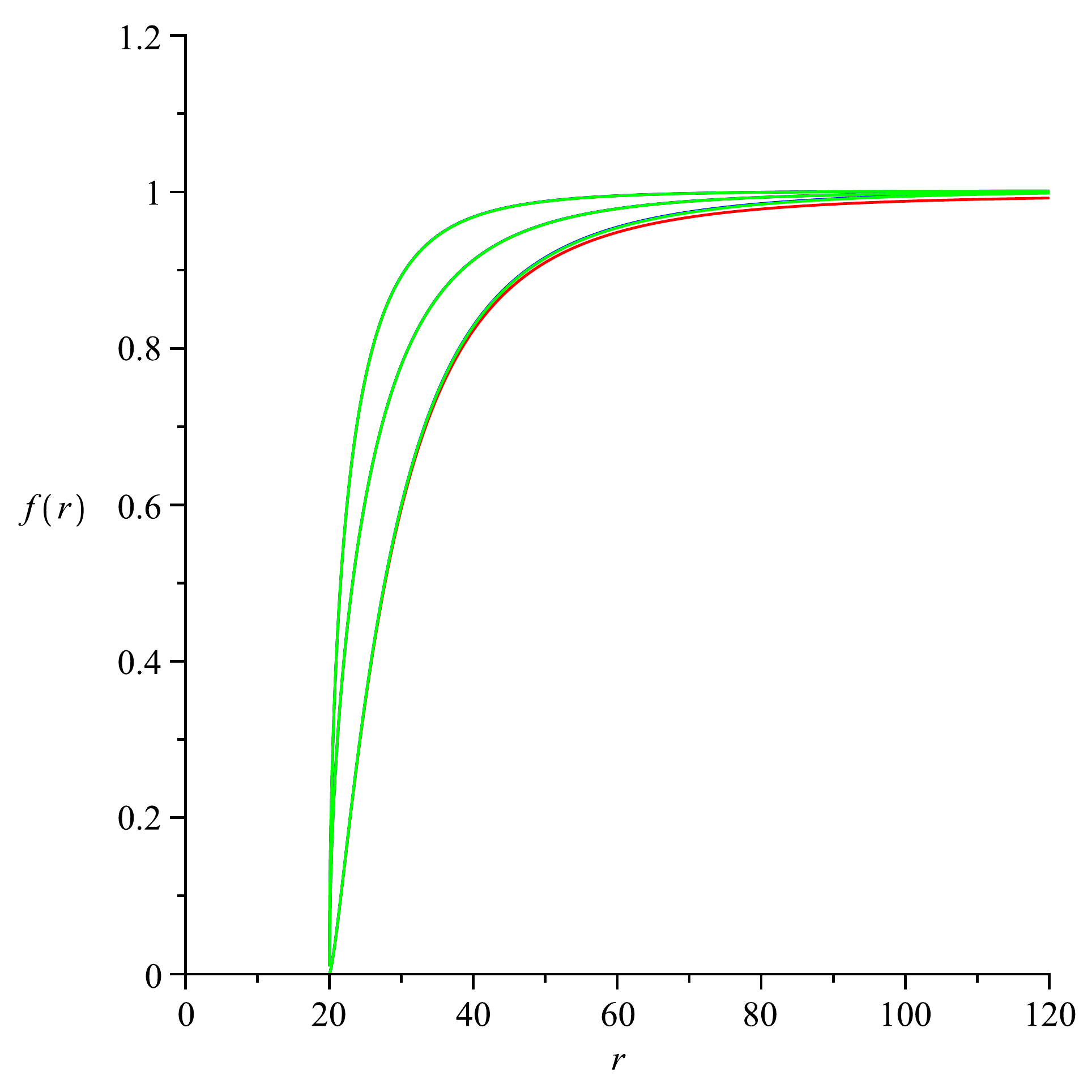}}
\caption{Metric functions $f(r)$ for $r_{0}=20$, $z=2$ and all
$k$. Left) in 5 dimensions and $Q=0,\,30000,\,46600$
from top to bottom [$Q_{c}\approx 46600$]. Right)  in 4
dimensions and $Q=0,\,1100,\,1780$ from top to bottom
[$Q_{c}\approx 1788$].  In each case $k=0$ is red, $k=1$ is blue and $k=-1$ is green; these different cases are just barely distinguishable for such large values
of $r$ .} \label{fr20n6z2}
\end{figure}

\begin{figure}[tbp]
\centering
{\includegraphics[width=.4\textwidth]{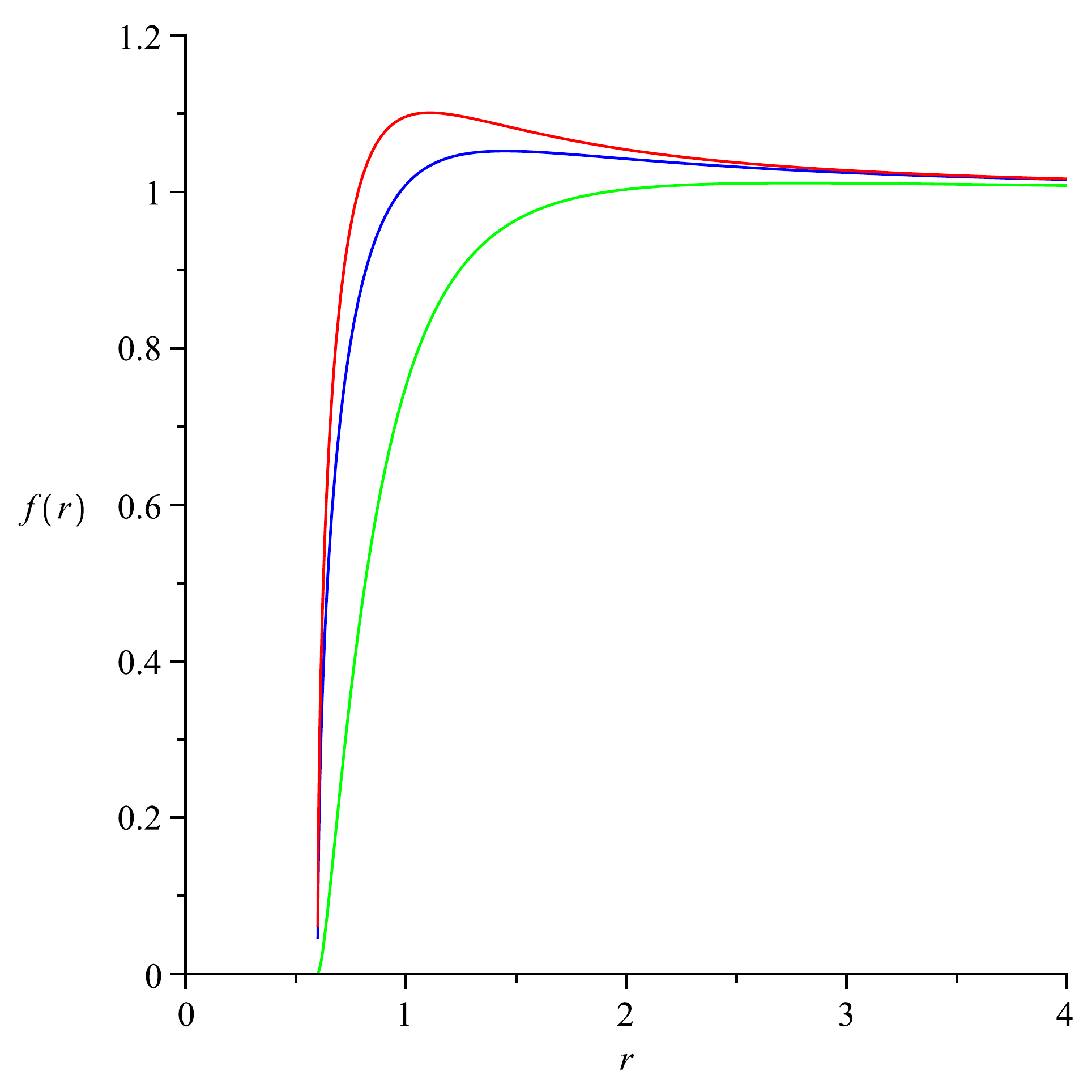}}\qquad\qquad {%
\includegraphics[width=.4\textwidth]{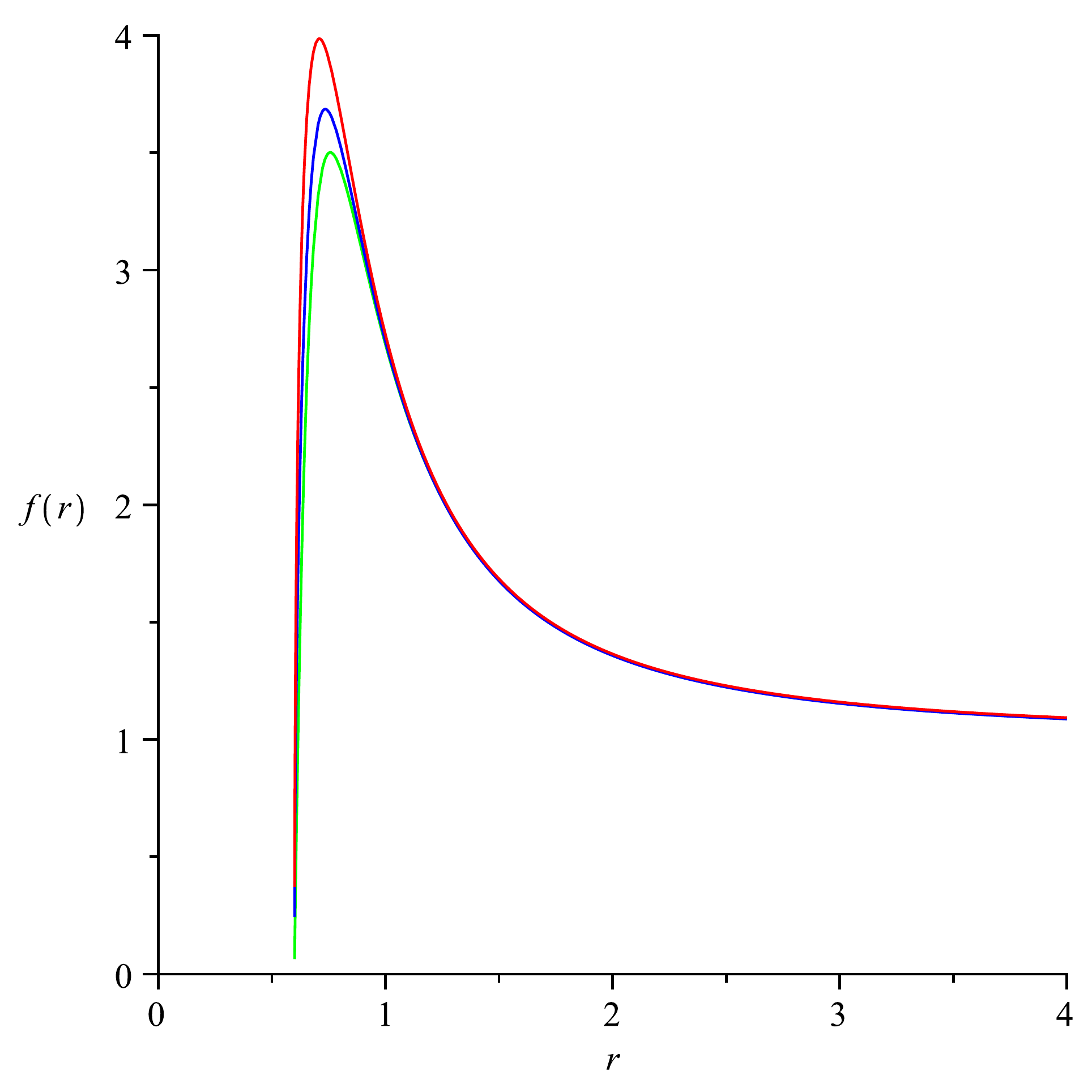}}
\caption{Metric functions $f(r)$ for different dimensions. Left) 4-dimensions with
$r_{0}=0.6$ for
zero mode and $k=1$ for $Q=0,\,1,\,2$ from top to bottom, respectively [$%
Q_{c}\approx$ 2]. Right) 7-dimensions with $r_{0}=0.6$ for zero mode and $%
k=-1 $ for $Q=0,\,0.3,\,0.41$ from up to down, respectively [$Q_{c}\approx$%
0.417]. Increasing values of $Q$ are coloured red, blue, and green respectively in each case.} \label{frQ5}
\end{figure}

We now consider how the Maxwell gauge field changes the solutions.
First, the requirement \cite{Zingg2} that the field
strength $(r^z\kappa) ^{\prime }$ of the massless gauge field
vanishes as $r$ goes to infinity implies $z\leq (n-1)$.
For the zero mode case ($z=n-1$), we find
that if $Q\ll Q_{c}$ these solutions are not significantly
different from uncharged ones. However distinctions start to
appear as the Maxwell charge $Q$ becomes comparable to $Q_{c}$.
Indeed in each dimension, for large black holes both $Q_c$ and the
metric/gauge functions are almost indistinguishable for differing
values of $k$ and the same charge, as shown in Fig.
(\ref{fr20n6z2}). This is not the case for small black holes. Here
$Q_c$ significantly depends on $k$ and the solutions are
distinguishable for differing $k$. However in general, as $Q$
increases the solutions either approach unity at large $r$ less
rapidly or become less sharply peaked for intermediate values of
$r$ in situations where such peaks exist. For example, three
different solutions for small black holes of the same radius but
different Maxwell charges are illustrated in Fig. (\ref{frQ5}).
The left-hand side of the diagram is for $k=1$ in 4-dimensions and
the right-hand side is for $k=-1$ in 7-dimensions.

\begin{figure}[tbp]
\centering {\includegraphics[width=.3\textwidth]{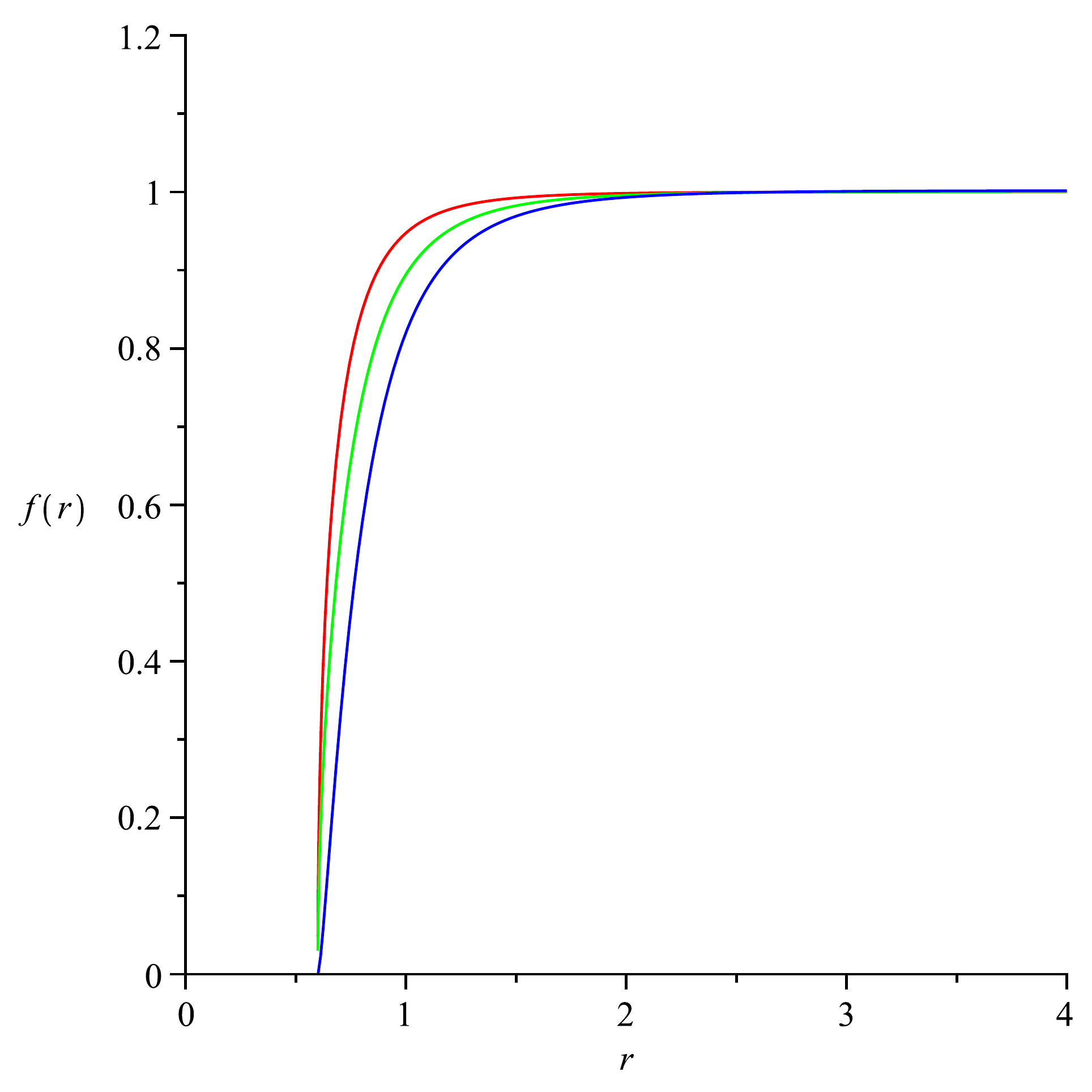}}\,
{\includegraphics[width=.3\textwidth]{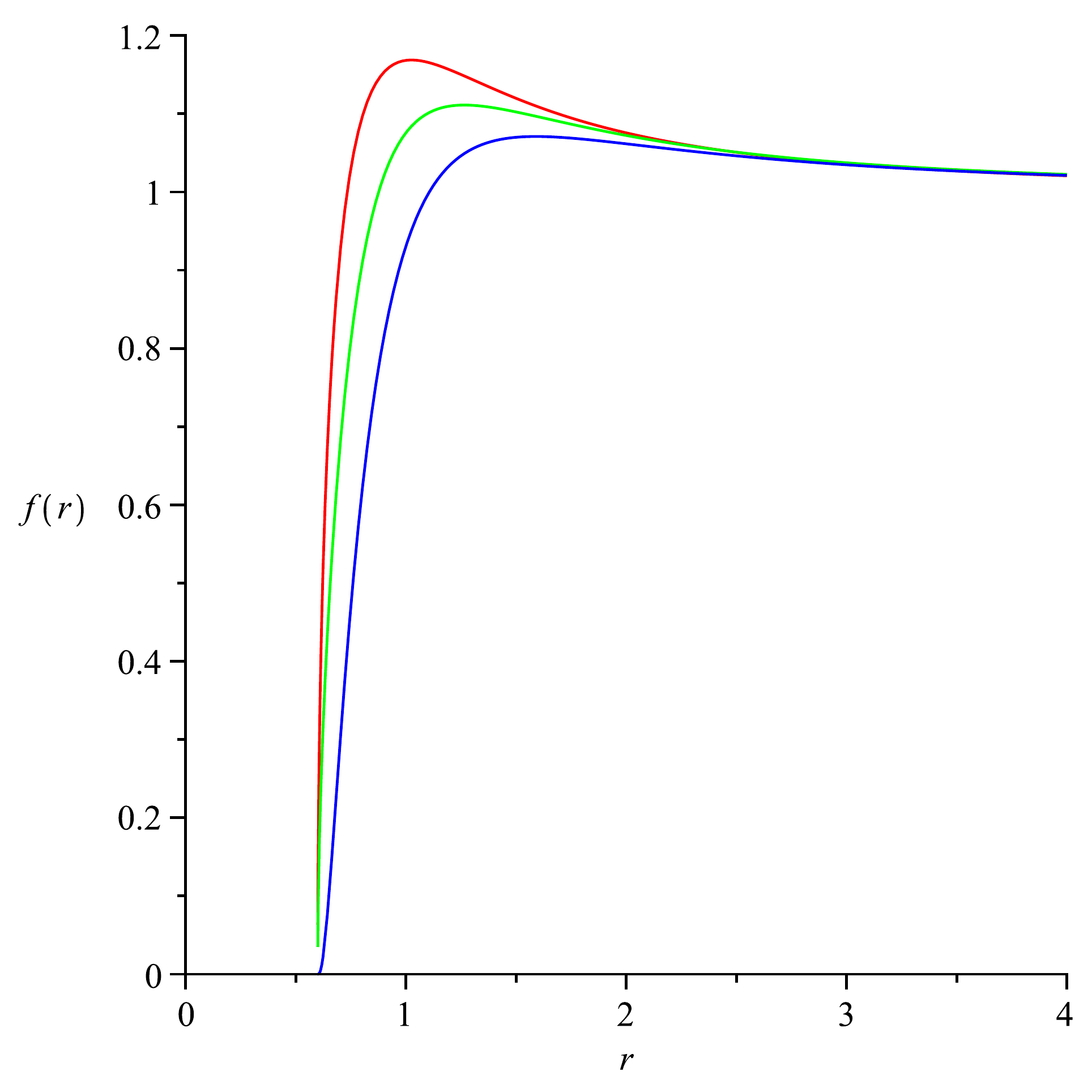}}\,
{\includegraphics[width=.3\textwidth]{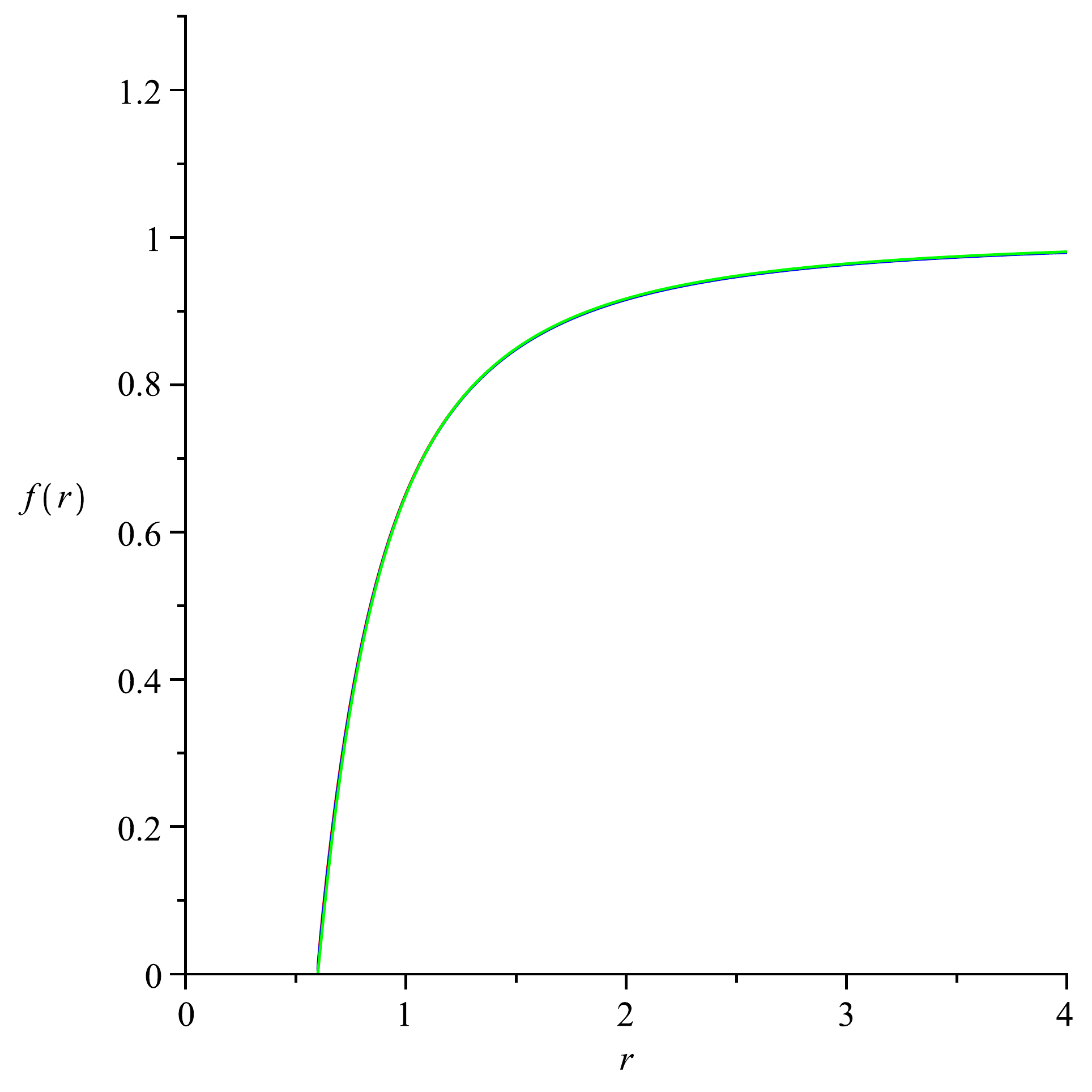}}
\caption{metric functions $f(r)$ in 5 dimensions for $r_{0}=0.6$
and $z=2$ for: Left)  $k=0$ and $Q=0,\,0.8,\,1.259$
[$Q_{c}\approx$ 1.259], Middle)  $k=1$ and $Q=0,\,1.1,\,1.77$
[$Q_{c}\approx$ 1.77],Right) $k=-1$ and $Q=0,\,0.1,\,0.176$
[$Q_{c}\approx$ 0.176] from top to bottom, respectively.
Increasing values of $Q$ are coloured red, blue, and green
respectively in each case.} \label{frn6z2}
\end{figure}
If $z<n-1$,  for a fixed $r_{0}$ there is a family of solutions
depending on fall-off rate. Consequently to investigate how
the Maxwell charge modifies the solutions by demanding the same
fall-off rate (for a given value of $k$) we search for possible charged black holes.
Numerical exploration reveals that with a fixed $r_{0}$, we cannot
find a solution if $Q>Q_{c}$, where $Q_{c}$ is given by eq. (\ref
{Qc}). Fig. (\ref{frn6z2}) shows $f(r)$ for small black holes in 5
dimensions with $z=2$ and $k=0,\, 1,\, -1$ for different $Q$.

Corresponding plots can be constructed for $g(r)$ and $h(r)$; we have not reproduced them  here.

\section{Thermal behavior}

The temperature of Lifshitz black holes is easily evaluated using
standard Wick rotation methods, yielding the result
\begin{equation}  \label{temp}
T=\frac{f_{0}r_{0}^{z+1}}{4\pi g_{0}},
\end{equation}
where $g_{0}$ is given by (\ref{g0}) and $f_{0}$ is determine so
that the metric has asymptotic behavior given in equation
(\ref{asmet}). The temperature clearly depends on the spacetime
dimensionality $(n+1)$, the
critical exponent $z$, the topological parameter $k$, the horizon radius $%
r_{0}$, and the Maxwell charge $Q$.

\begin{figure}[tbp]
\centering
{\label{tempQ0km}\includegraphics[width=0.3\textwidth]{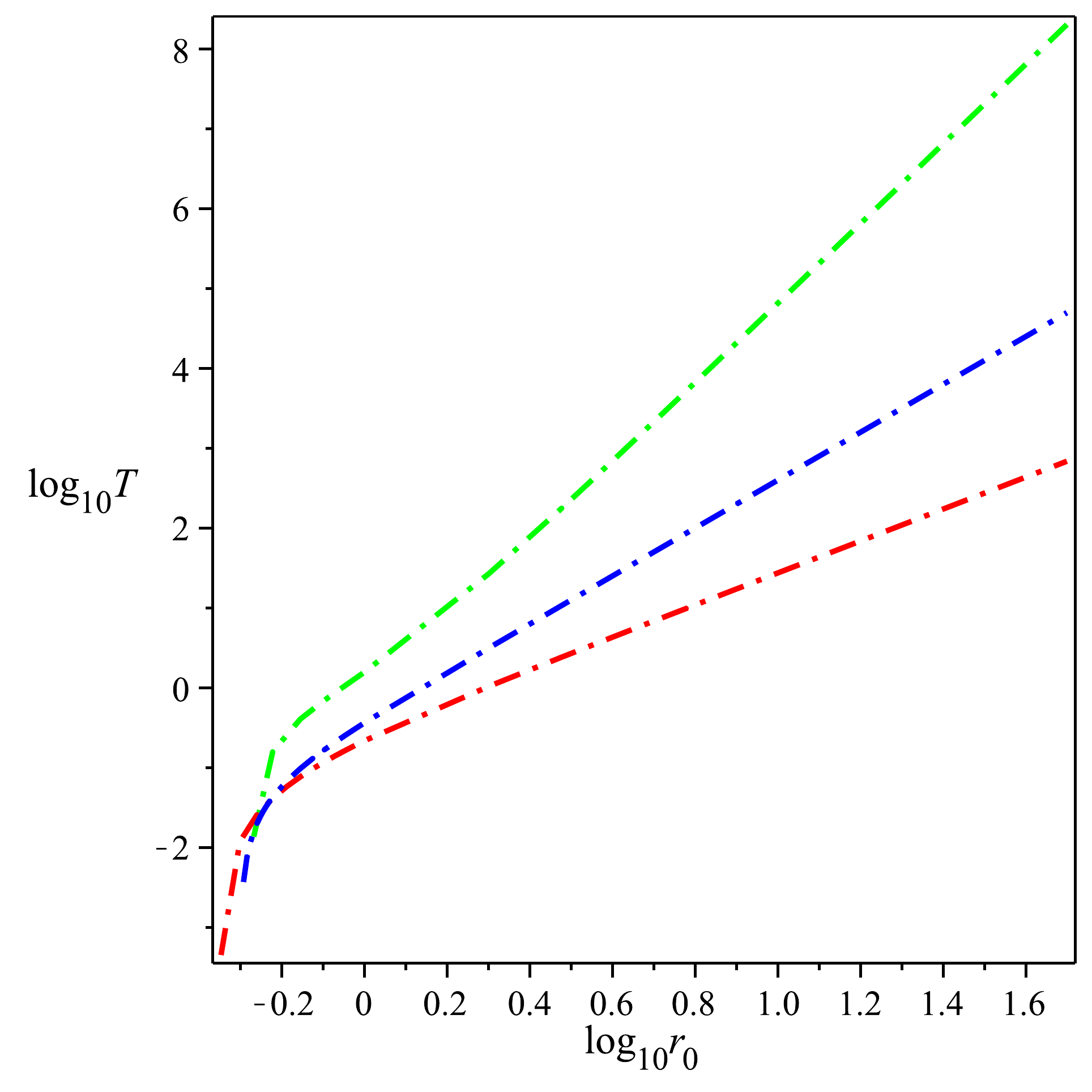}} {%
\label{tempQ0k0}\includegraphics[width=0.3\textwidth]{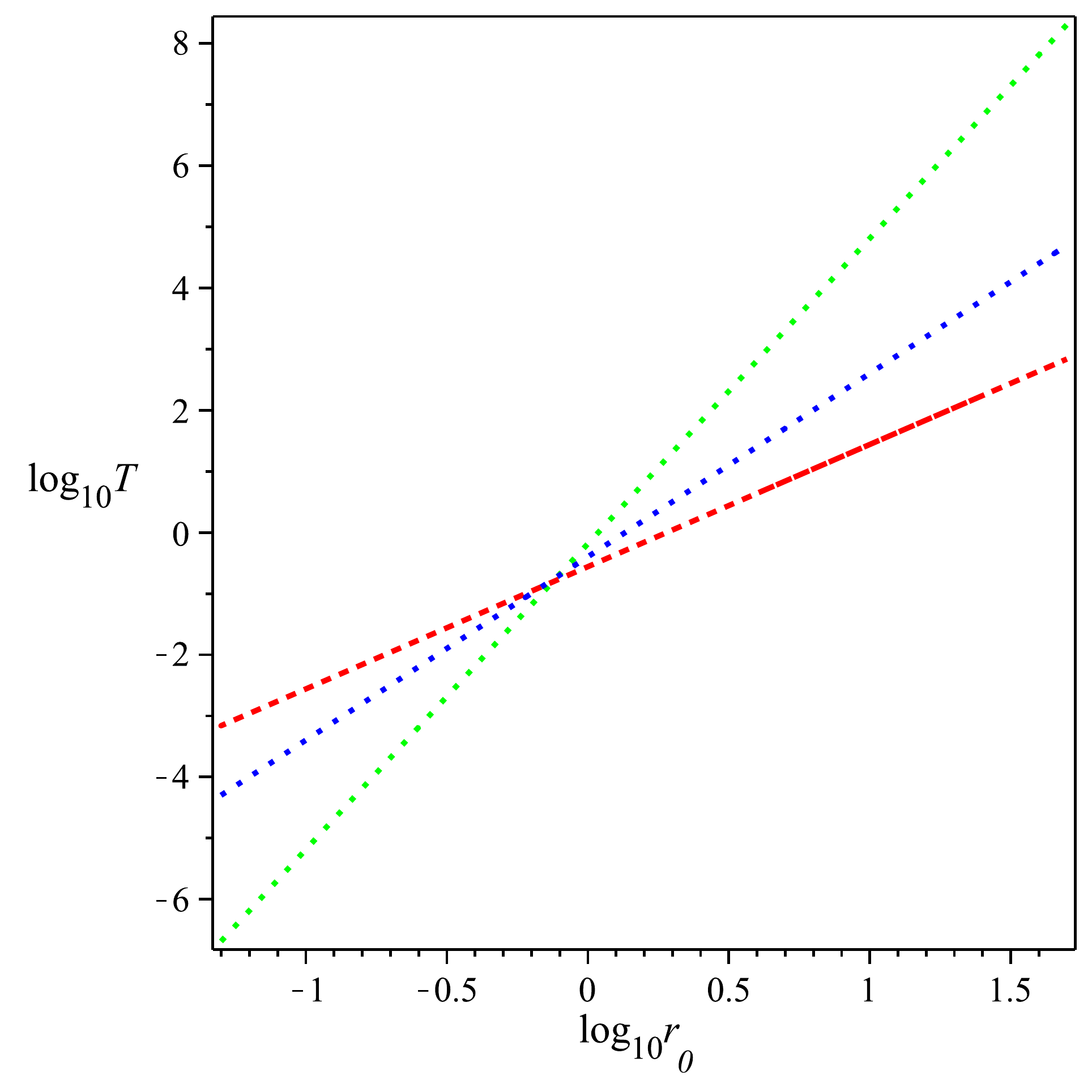}} {%
\label{tempQ0kp}\includegraphics[width=0.3\textwidth]{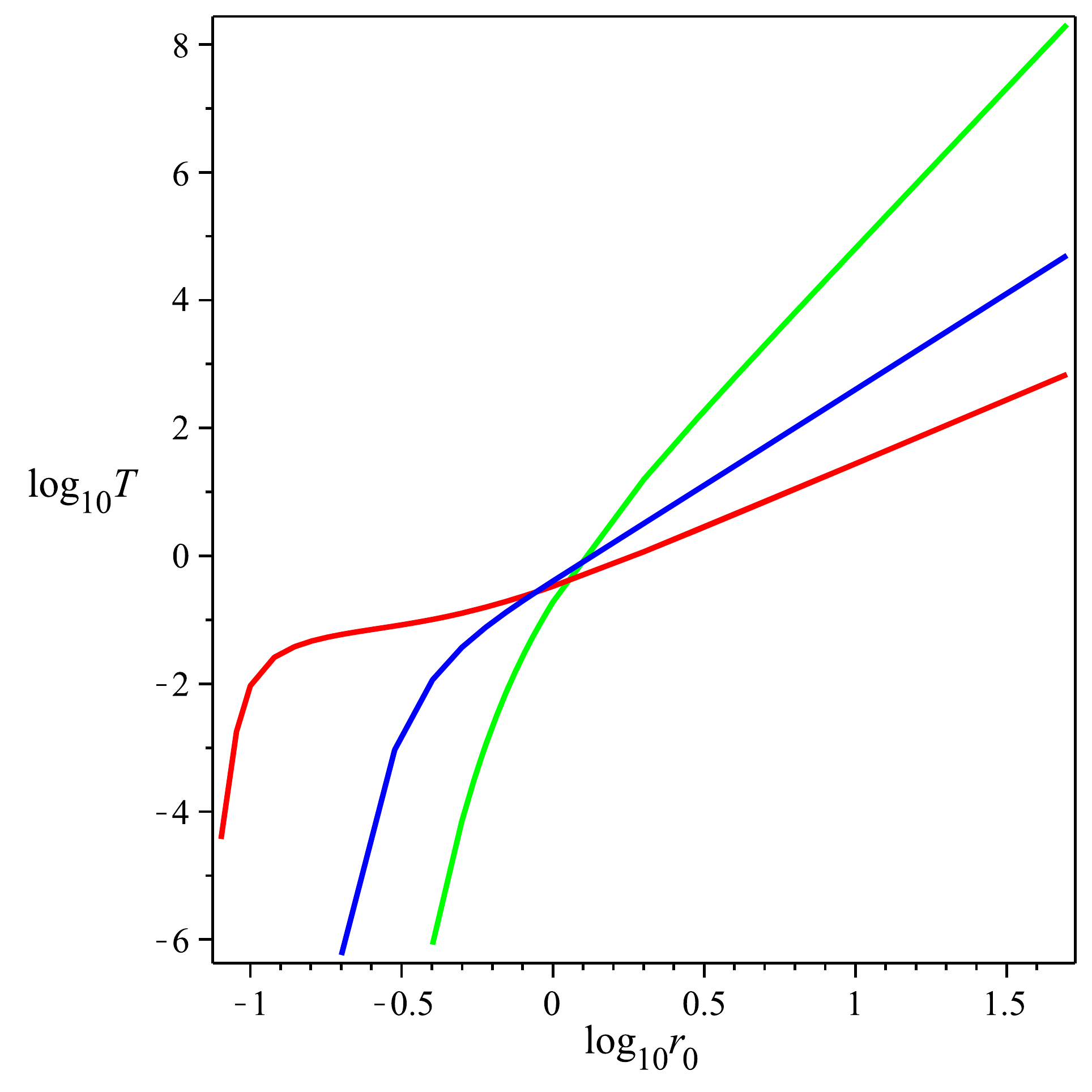}}
\caption{Temperature vs $r_{0}$ with $Q=0$ for zero modes in
4(red), 5(blue) and 7(green)-dimensions for: $k=-1$ (dash-dots),
$k=0$ (dots) and $k=1$ (solids).} \label{tempr0}
\end{figure}

The behavior of temperature versus $r_{0}$ for uncharged solutions
$Q=0$ and
zero modes $z=n-1$ in different dimensions is depicted in Fig. (\ref{tempr0}%
) for $k=-1, 0$ and 1. For all topologies we see that the
temperature increases with dimensionality for large black holes,
whereas it decreases with dimensionality for small black holes.
This effect is most pronounced for $k=1$, and least so for $k=-1$.
In each dimension the temperature decreases as the horizon radius
shrinks, i.e. small black holes are colder than large ones. For
$k=-1$, there is  a lower bound on the radius of the black hole.

For $Q\neq 0$, since the metric functions $f(r)$ and $g(r)$ are
not altered significantly when $Q$ is small compared to its upper
bound  $Q_c$ no significant change in the behaviour of the
temperature is expected. This upper bound value increases with
increasing horizon radius. Indeed, with a fixed horizon the
temperature goes to zero as $Q\rightarrow Q_c$. We therefore
consider the
behaviour of temperature as a function of $Q$ for fixed horizon radius $%
r_{0} $.

\begin{figure}[tbp]
\centering
{\includegraphics[width=0.3\textwidth]{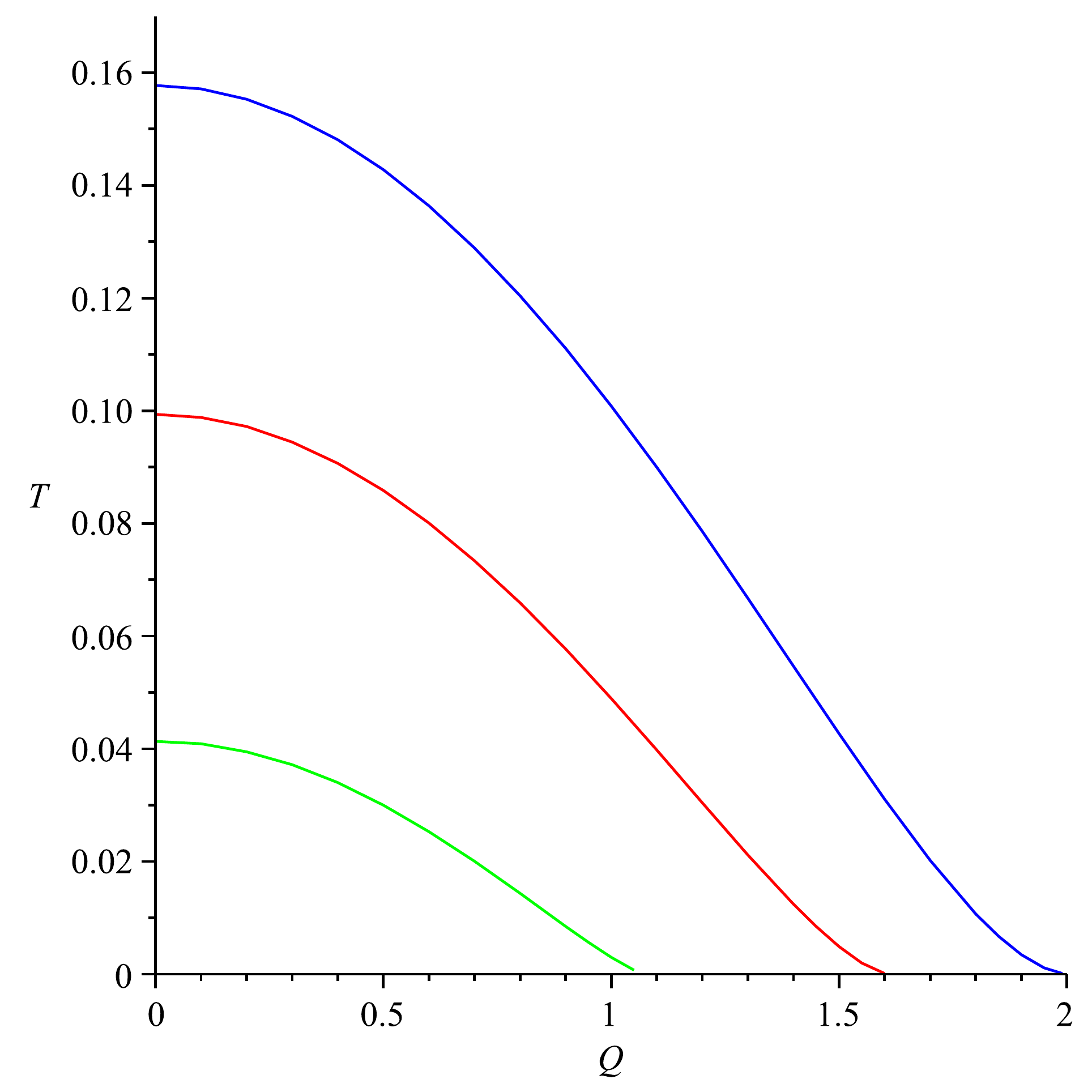}\qquad\qquad} {%
\includegraphics[width=0.3\textwidth]{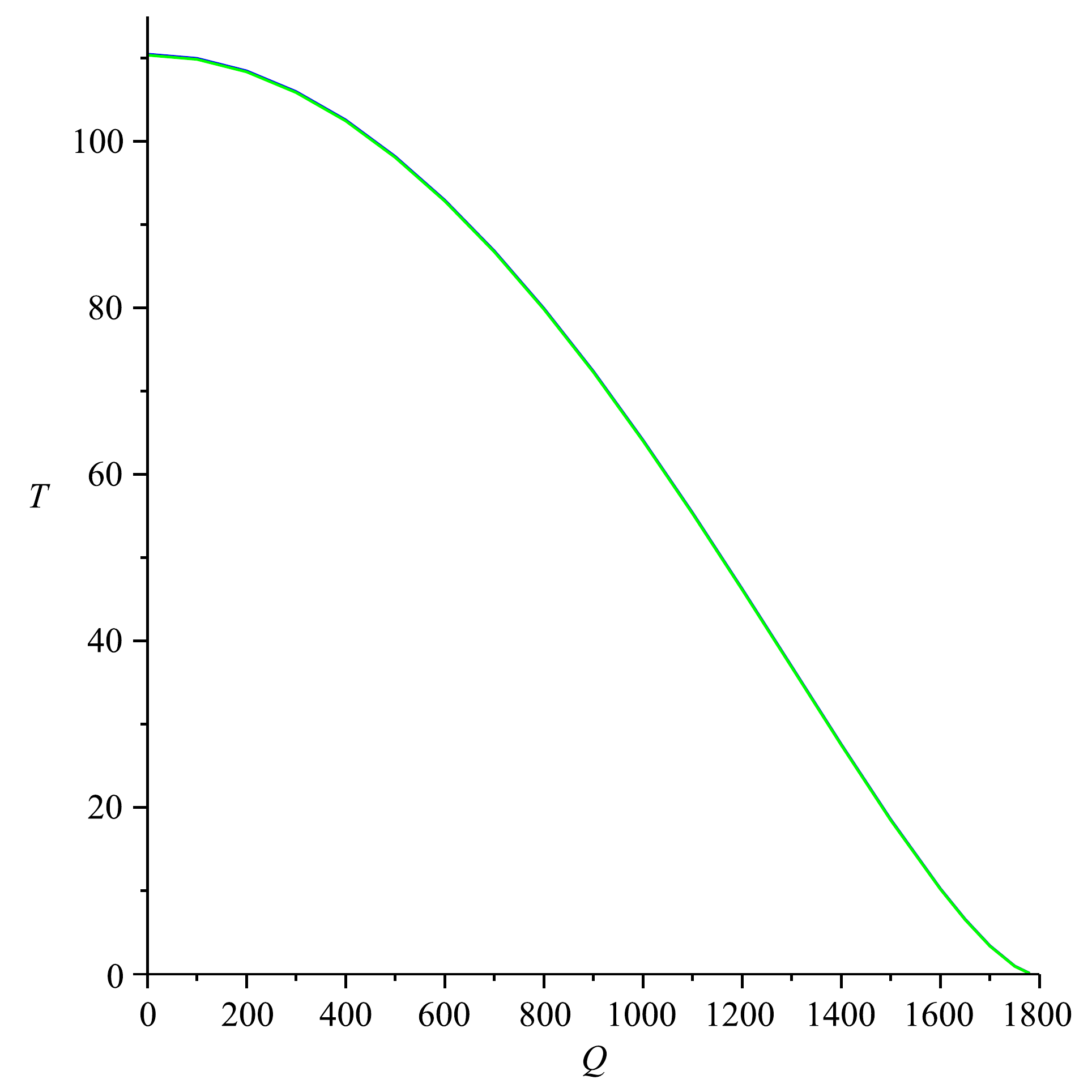}}
\caption{Temperature vs $Q$ in 4-dimensions with $z=2$ for: $r_{0}=0.6$ for $%
k=-1$ (green), $k=0$ (red) and $k=1$ (blue) on the left and
$r_{0}=20$ on the right; in the latter case the curves are almost identical for
each $k$.} \label{tempQ4}
\end{figure}

\begin{figure}[tbp]
\centering
{\includegraphics[width=0.3\textwidth]{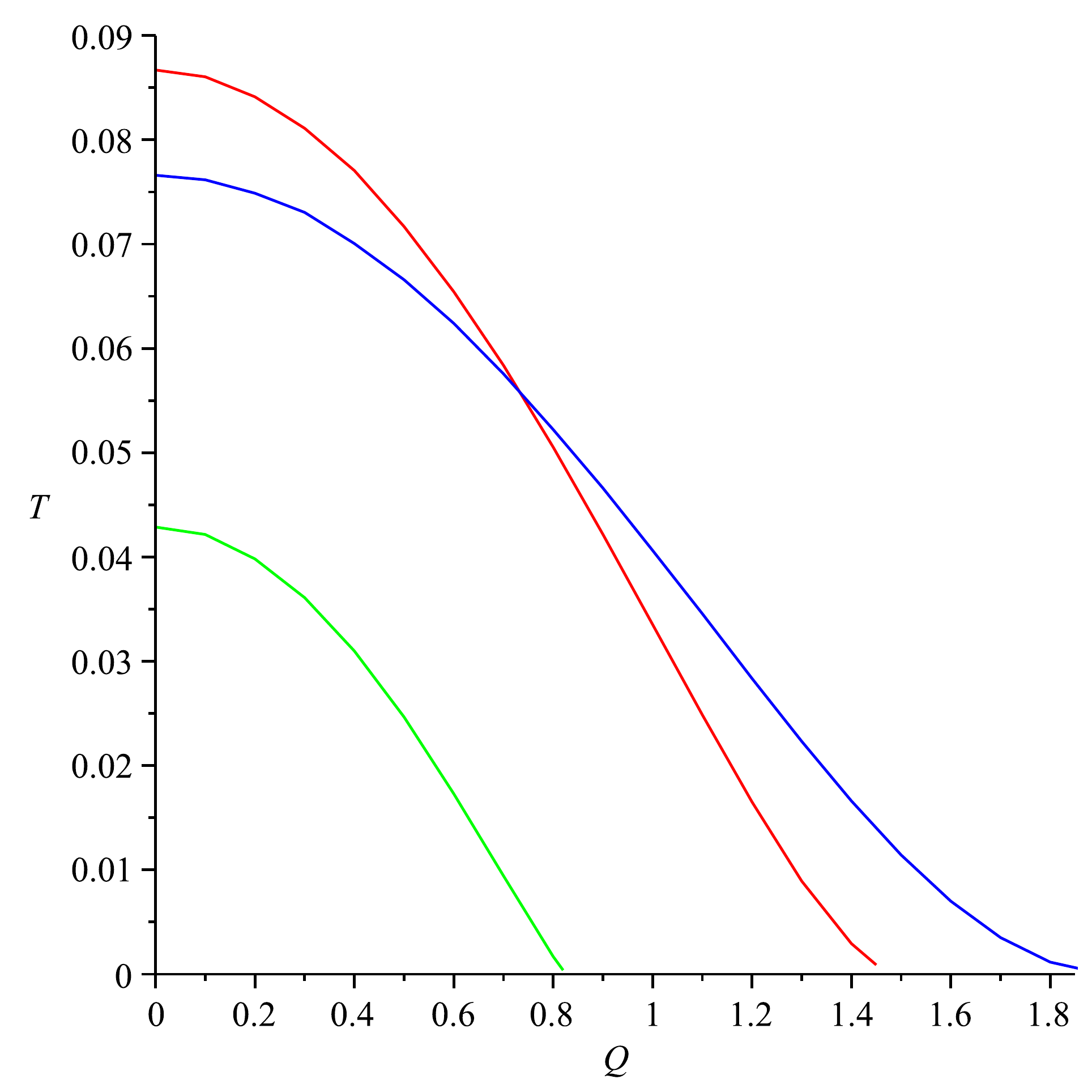}\qquad\qquad} {%
\includegraphics[width=0.3\textwidth]{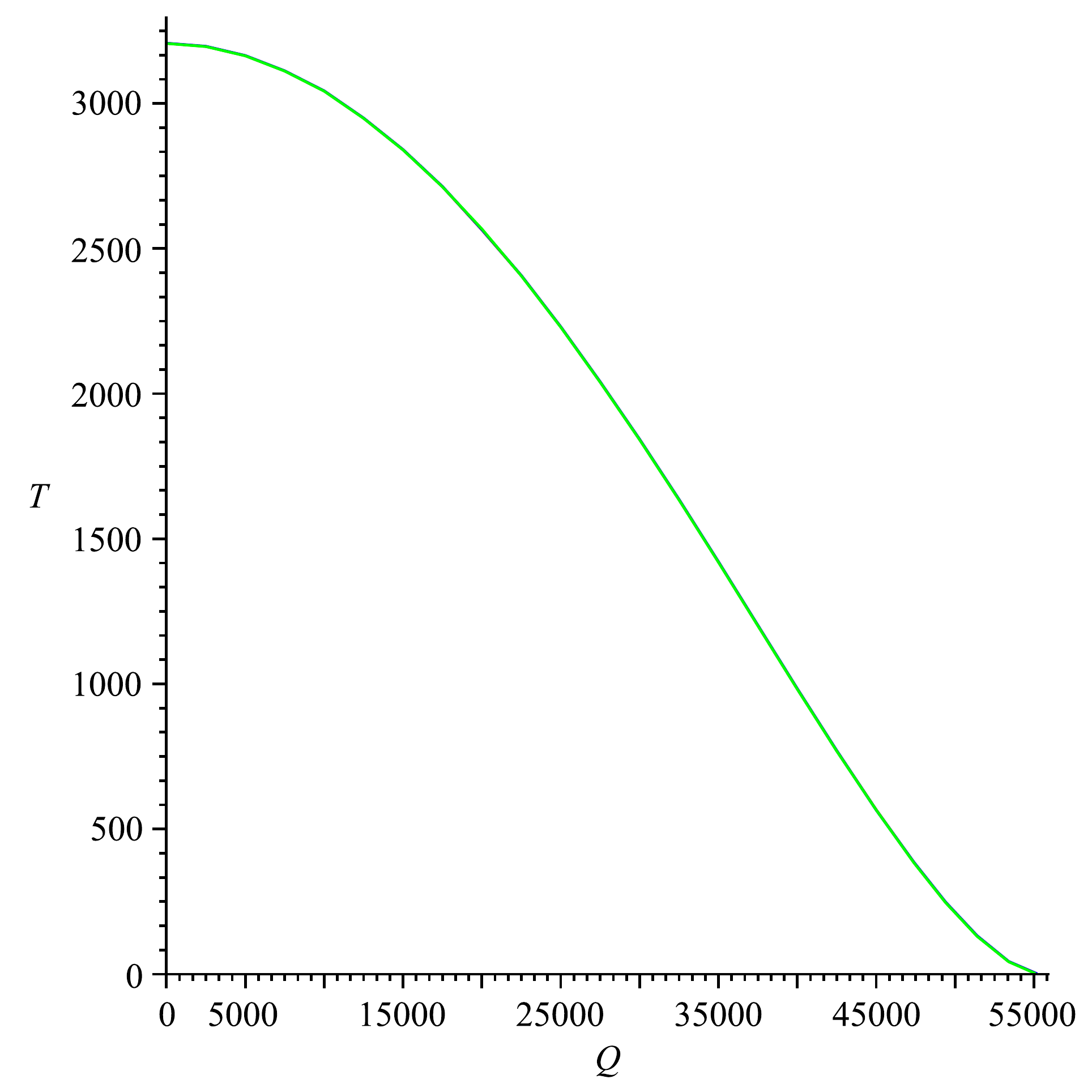}}
\caption{Temperature vs $Q$ in 5-dimensions with $z=3$ for: $r_{0}=0.6$ for $%
k=-1$ (dash-dot), $k=0$ (dot) and $k=1$ (solid) on the left and
$r_{0}=20$ for all $k$ on the right; in the latter case the curves are almost identical for
each $k$.} \label{tempQ5}
\end{figure}

Figure (\ref{tempQ4}) illustrates the situation in 4-dimensions
for both small ($r_0=0.6$) and large ($r_0=20$) black holes. For
all topologies the temperature decreases with increasing $Q$ as
expected, with temperature being largest for the spherical case
and smallest for the hyperbolic one. For large black holes this
latter effect is almost completely indistinguishable, whereas for
small black holes it is quite pronounced.
Furthermore, this effect does not hold as dimensionality increases. In Fig. (%
\ref{tempQ5}) we plot temperature versus $Q$ in 5-dimensions for
both small and large black holes. In the former case we see that
for sufficiently large $Q$ flat topologies yield hotter black
holes than spherical ones. In the latter case the distinctions
between topologies are very small -- they exhibit nearly the same
temperature for different $k$, a feature we find true in all
dimensions we have checked. In any dimension the temperature
approaches zero as the Maxwell charge approaches its extremal
value.


\section{Wilson loop}
In 4 dimensions the action for the dual theory is conjectured to
be
\begin{equation}
S=\int dtd^{2}\mathbf{x}\left( \dot{\phi}^{2}-K(\nabla ^{2}\phi
)^{z}\right) \label{bdyact}
\end{equation}
and if we write $\nabla ^{2}\phi =\vec{\nabla}\times \vec{E}$, where $%
E_{j}=\varepsilon _{jk}\nabla ^{k}\phi $, then the boundary action
(\ref {bdyact}) could be regarded as a gauge theory in (2+1)
dimensions with a dimensionless coupling constant \cite{Daniel,
Vishwanath}. We introduce Wilson loops by joining charged
particles on the boundary that are connected together in the bulk
via a string.   These loops contain information about the force
acting between particles charged under the gauge fields in the
dual theory.

The Euclidean action of this string for a rectangular Wilson loop
is the same for all values of $k$ and is given by
\cite{Vishwanath,Maldacena}
\begin{equation}
\mathcal{S}=\frac{1}{2\pi \alpha ^{\prime }}\int dtd\tau \sqrt{%
det[g_{AB}\partial _{\mu }X^{A}\partial _{\nu
}X^{B}]}=\frac{\triangle \ell
^{2}}{2\pi \alpha ^{\prime }}\int d\theta \sqrt{%
f^{2}r^{2z+2}+f^{2}g^{2}r^{2z-2}\left( \frac{dr}{d\theta }\right)
^{2}}
\end{equation}
taking $\sigma =\theta $ and $i\tau =t$ in the static gauge, with
Euclidean time interval $\triangle $.

\begin{figure}[tbp]
\centering
{\includegraphics[width=.4\textwidth]{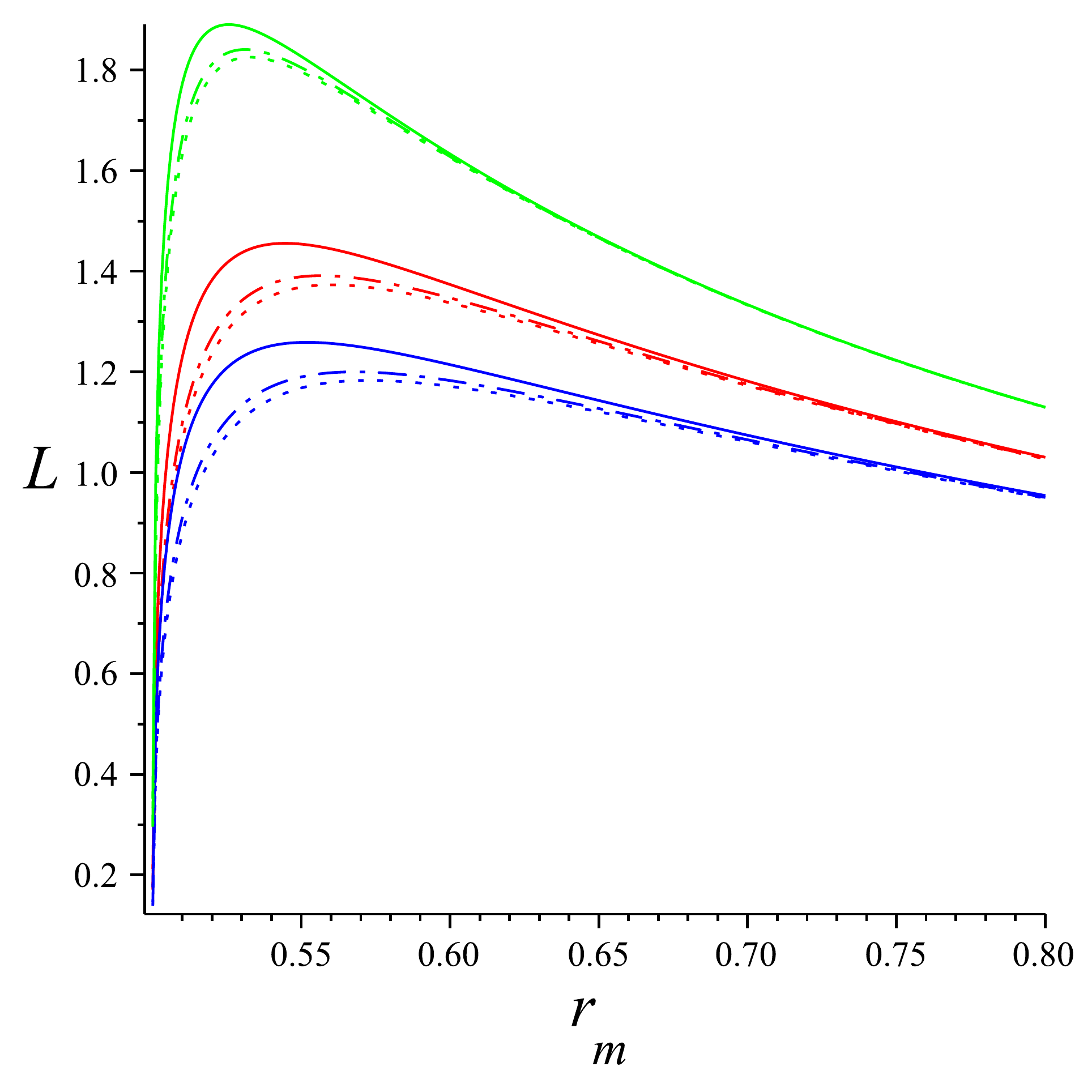}}
\caption{Boundary length $L$ vs $r_{m}$ in 4-dimensions with $z=2$ and $%
r_{0}=0.5$ for $k=1$ (blue), $k=0$ (red) and $k=-1$ (green) while
in each case dot is for $Q=0$, dash-dot for $Q=Q_{c}/3$ and solid
for $Q=2Q_{c}/3$.} \label{wloop}
\end{figure}

If one extremizes the action one may obtains a constant of the
motion as
\begin{equation}
\frac{f^{2}r^{2z+2}}{\sqrt{f^{2}r^{2z+2}+f^{2}g^{2}r^{2z-2}\left(\frac{dr}{%
d\theta}\right)^{2}}}=f(r_{m})r_{m}^{z+1}
\end{equation}
where $r_{m}>r_{0}$ is the location of the midpoint of the string for which $%
\frac{dr}{d\theta}\mid_{r_{m}}=0$. From the above expression one
may get the boundary length as
\begin{equation}
L=\int d\theta=2\int_{r_{m}}^{\infty}\frac{dr}{r^{2}}\frac{g}{\sqrt{\left(%
\frac{f}{f_{m}}\right)^{2}\left(\frac{r}{r_{m}}\right)^{2z+2}-1}}
\end{equation}
and the regularized potential energy between the two particles
\begin{equation}\label{pot}
V=\frac{\mathcal{S}}{\triangle\ell}=\frac{\ell}{2\pi\alpha^{\prime}}%
\left(2\int_{r_{m}}^{\infty}dr\frac{r^{z-1}fg} {\sqrt{1-\left(\frac{f}{f_{m}}%
\right)^{2}\left(\frac{r_{m}}{r}\right)^{2z+2}}}-2\int_{r_{0}}^{%
\infty}drr^{z-1}fg\right)
\end{equation}
where $f_{m}=f(r_{m})$.

We plot in Fig. (\ref{wloop}) how the behaviour of the Wilson loop
for the neutral case \cite{Daniel,Robb} is modified for nonzero
$Q$. For a given $r_m$ we find that the boundary length increases and
the potential between objects in the dual theory decreases (Fig. (\ref{pot})) with increasing
Maxwell charge.

\begin{figure}[tbp]
{\centering
{\label{tempQ0km}\includegraphics[width=0.3\textwidth]{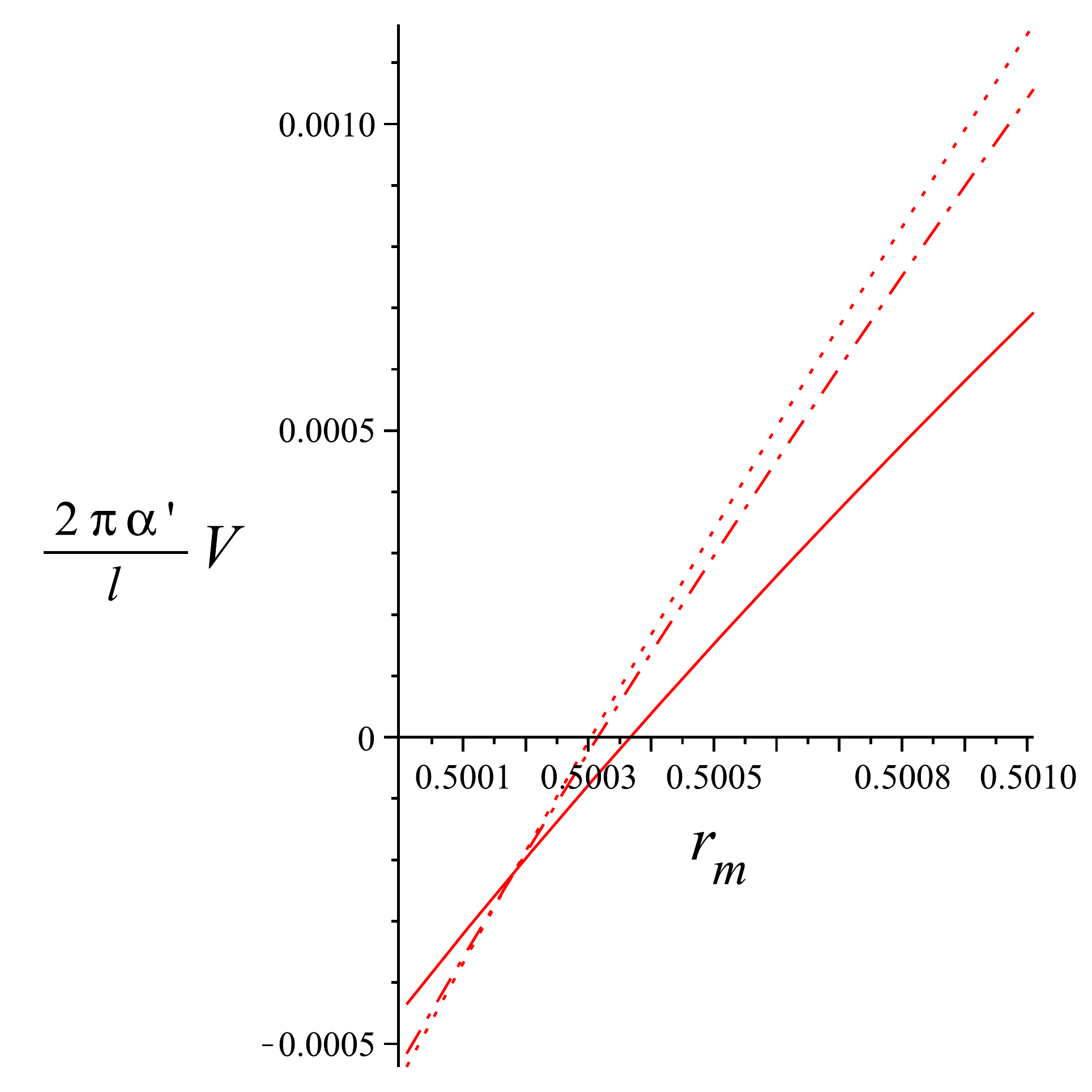}} {\label%
{tempQ0k0}\includegraphics[width=0.3\textwidth]{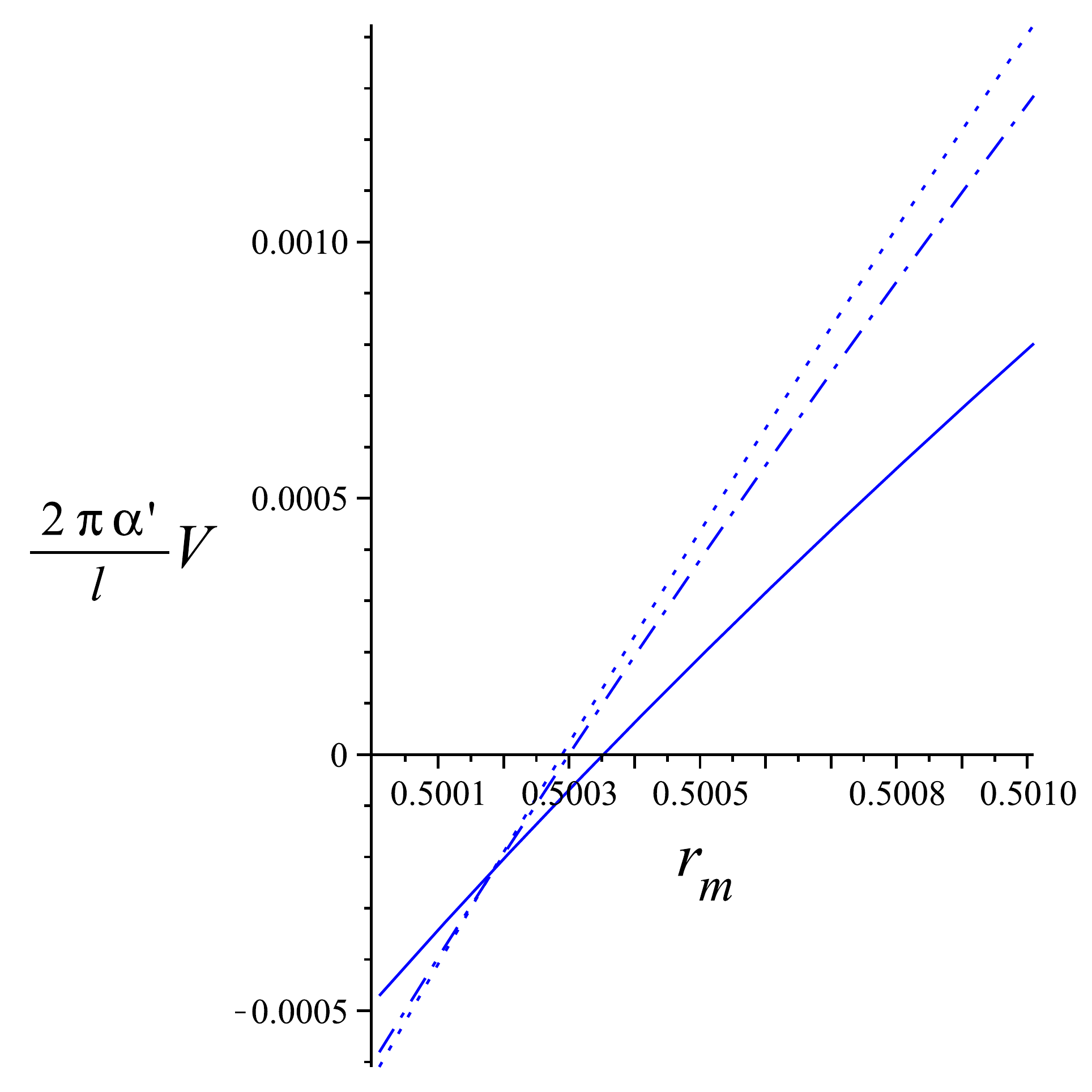}} {\label%
{tempQ0kp}\includegraphics[width=0.3\textwidth]{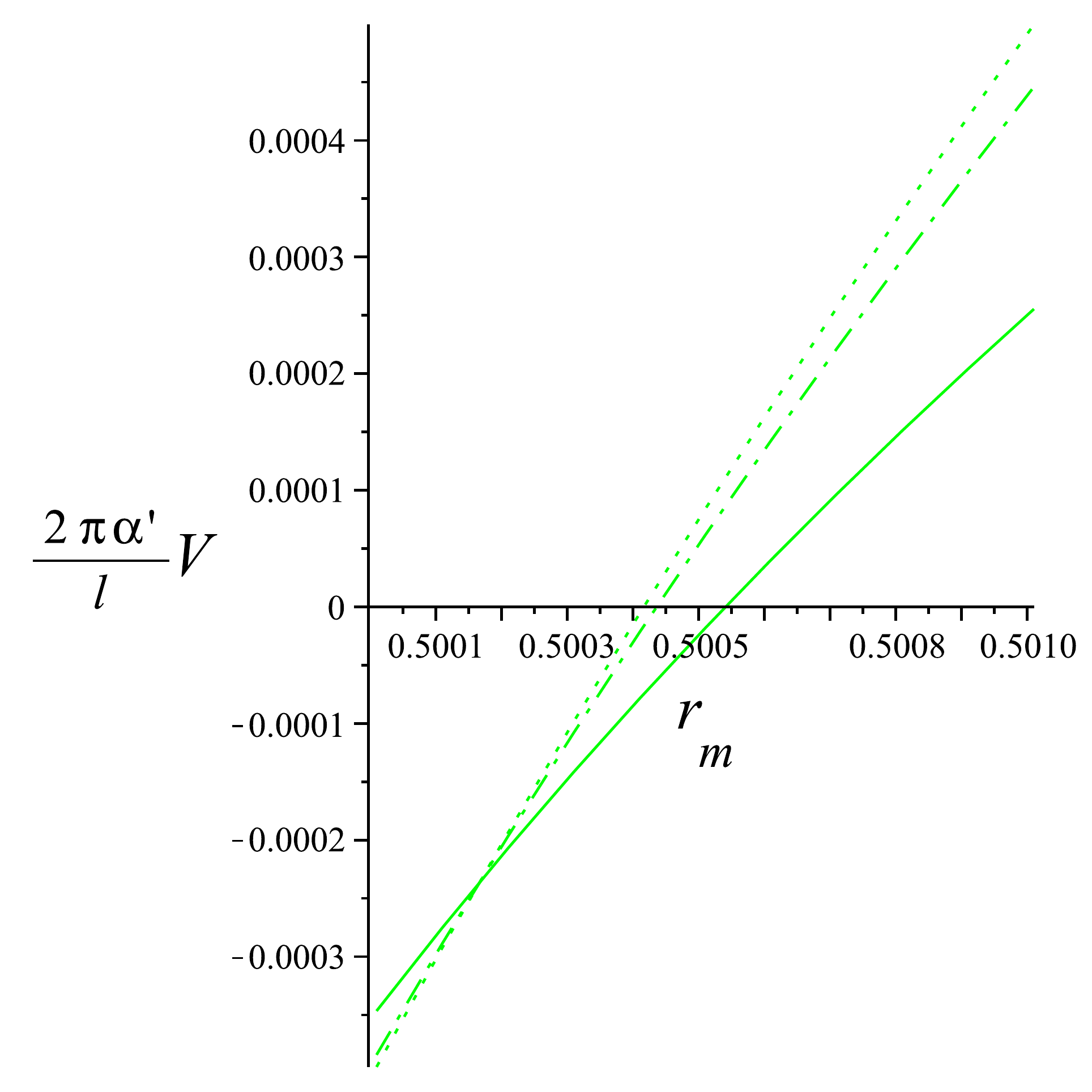}} }
\caption{Potential $V$ between two particles as a function of
string midpoint $r_{m}$ in 4-dimensions with $z=2$ and $r_{0}=0.5$
for $k=0$ (red),
$k=1$ (blue) and $k=-1$ (green) for $Q=0$ (dot), $Q=Q_{c}/3$ (dash-dot) and $%
Q=2Q_{c}/3$ (solid).} \label{pot}
\end{figure}

\section{Conserved Charge}

To explore the conserved charge we redefine the metric and gauge
fields as follows while we have chosen $\ell =1$ for simplicity
through the rest of the paper:
\begin{eqnarray}
&&ds^{2}=-e^{2F(r)}dt^{2}+e^{2G(r)}dr^{2}+e^{2R(r)}d\mathbf{x}^{2},
\nonumber \\
&&A_{t}=e^{K(r)},\qquad B_{t}=qe^{J(r)},\qquad H_{tr}=qe^{H(r)}.
\label{newansatz}
\end{eqnarray}
where new fields $F,\,G,\,R,\,K,\,J,\,H$ can be expressed in terms
of the old fields as:
\begin{eqnarray}
&&F(r)=\ln (r^{z}f),\quad G(r)=\ln (g/r),\quad R(r)=\ln {r}, \\
&&H(r)=\ln (zr^{z-1}ghf),\quad J(r)=\ln (r^{z}jf),\quad
K(r)=\ln(r^{z}\kappa).
\end{eqnarray}

If one inserts the ansatz (\ref{newansatz}) into the action
(\ref{action}),
one obtains the one dimensional lagrangian $\mathcal{L}_{1D}=\mathcal{L}%
_{1g}+\mathcal{L}_{1m}$ with
\begin{eqnarray}
\mathcal{L}_{1g} &=&(n-1)\left[ -\frac{2\Lambda
e^{2G}}{n-1}+2F^{\prime
}R^{\prime }+(n-2)R^{\prime \prime }\right] e^{F-G+(n-1)R},  \nonumber \\
\mathcal{L}_{1m} &=&\frac{1}{2}e^{-F+G+(n-1)R+2J}\left[
q^{2}(C+J^{\prime 2}e^{-2G})+K^{\prime 2}e^{2K-2J-2G}\right] .
\end{eqnarray}
where we have used eq. (\ref{EqB}) to write $H(r)$ in terms of
$J(r)$. Applying the same method described in \cite{HR} reveals
that
\begin{eqnarray}
\mathcal{C} &\equiv &2(F^{\prime }-R^{\prime
})e^{F-G+(n-1)R}-(q^{2}J^{\prime }e^{2J}-K^{\prime
}e^{2K})e^{-F-G+(n-1)R}
\nonumber \\
&=&\frac{r^{n+z-1}}{g}\left[ 2(z-1)f+2rf^{\prime
}-q^{2}j(zfj+rfj^{\prime }+rjf^{\prime })\right]-Qr^z\kappa,
\label{C0}
\end{eqnarray}
is conserved along the radial coordinate $r$ and
\begin{equation}
\kappa(r)=\frac{1}{r^z}\left[Q\int^r
\frac{fg}{r^{n-z}}dr+D\right],\label{k1r}
\end{equation}
where $D$ is an integration constant,  which is chosen such that $\kappa(r)$ vanishes at the horizon, that is
\begin{equation}
D=-Q\int^{r_0} \frac{fg}{r^{n-z}}dr. \label{DD}
\end{equation}
 We shall evaluate $\mathcal{C}$ both at the horizon
radius and infinity. Now using (\ref{k1r}) and the large $r$
expansions (\ref{glr}), (\ref{hlr}), (\ref{jlr}) and (\ref{flr})
conserved charge for $z\neq n-1$ is given by:
\begin{equation}
\mathcal{C}=\frac{2(z-1)(n-z-1)(z+n-1)}{(n-1)^{2}}C_{1}-QD.
\label{Ceqn1}
\end{equation}
and for $z=n-1$ is:
\begin{equation}
\mathcal{C}=\frac{Q^2}{2(n-1)}-\frac{2C_1}{n-1}, \label{Ceqn2}
\end{equation}
Using the near horizon expansions (\ref{Eqnh}) we find that
\begin{equation}
\mathcal{C}=\frac{f_{0}}{g_{0}}r_{0}^{n+z}=16\pi
TS,\label{Chorizon}
\end{equation}
while we have considered massless field is zero at the horizon and
$S$ is the entropy of the black brane per unit volume which in
terms of the horizon radius is given by:
\begin{equation}
S=\frac{1}{4}r_{0}^{n-1}. \label{Entropy}
\end{equation}
We remark that the conserved quantity $\mathcal{C}$ for $z = 1$
with $g(r) = 1/f(r)$ reduces to
\begin{equation}
\mathcal{C}=2r^{n+1} f f^{\prime}+\frac{Q^2}{(n-2)
r^{n-2}},\label{consads}
\end{equation}
which is proportional to the mass of the asymptotically
AdS-Reissner-Nordstrom solution and gives the metric function as
\begin{equation}
f^2(r)=1-\frac{m}{r^{n}}+\frac{Q^2}{2(n-1)(n-2)
r^{2(n-1)}},\label{frads}
\end{equation}
which is the well-known asymptotically AdS Reissner-Nordstrom
solution with flat horizon.

\section{Finite Action and the Energy Density for Einstein-Maxwell-Lifshitz
Solutions}

In order to have a finite action we must add some boundary terms
to the original bulk action. For this purpose, we use the same
method proposed in \cite{Saremi} to make ($n+1$)-dimensional
Einstein-Maxwell-Lifshitz(EML)
action finite and well-defined. We consider $I=I_{\mathrm{bulk}}+I_{\mathrm{%
bdy}}$, where $I_{\mathrm{bulk}}$ is given in eq. (\ref{action}) and $I_{%
\mathrm{bdy}}$ is the boundary action which for the case of zero
curvature boundary can be written as:
\begin{equation}
I_{\mathrm{bdy}}=\frac{1}{8\pi}\int_{\partial\mathcal{M}}d^{n}x\sqrt{-h}%
\left[K-(n-1)+\frac{1}{2}f(B_{\alpha}B^{\alpha})\right]+\frac{\omega}{8\pi}%
\int_{\partial\mathcal{M}}d^{n}x\sqrt{-h}n^{\mu}F_{\mu\nu}
A^{\nu}+I_{\mathrm{deriv}},\label{actbdy}
\end{equation}
where in the first integral, which is added to cancel the
divergences from gravity and the Lifshitz field, the boundary
$\partial\mathcal{M}$ is the hypersurface at some constant $r$,
$h_{\alpha\beta}$ is the induced metric, $K$ is the trace of the
extrinsic curvature, $K_{\alpha\beta}=\nabla_{(\alpha}n_{\beta)}$
of the boundary (where the unit vector $n^{\mu}$ is orthogonal to
the boundary and outward-directed).

In the grand canonical ensemble we set $\delta A^{\mu}=0$ on the boundary and the
variational principle is well-defined only if $\omega=0$.  In the canonical ensemble
we hold the electric charge  (given by eq. (\ref{Qconserve}))
fixed, which is equivalent to setting
$\delta (\sqrt{-h}n^{\mu}F_{\mu\nu}) = 0$ on the boundary \cite{RobbRoss}.
In this latter instance the variational principle is well-defined
provided we select $\omega=\frac{1}{2}$.
 The last term  $I_{\mathrm{deriv}}$ in (\ref{actbdy}) is
a collection of terms involving derivatives of the boundary
fields, which could involve both the curvature tensor constructed
from the boundary metric and covariant derivatives of
$B_{\alpha}$. Since the boundary is flat and the fields are
constants for (\ref{asmet}), this term will not contribute to the
on-shell value of the action for the pure Lifshitz solution or its
first variation around the Lifshitz background and therefore we
ignore it henceforth. As explained in \cite{Saremi}, an arbitrary
function $f(B^{\alpha}B_{\alpha})$
is added to the action which is due to the fact that on the boundary $%
B_{\alpha}B^{\alpha}=-q^{2}$ is constant for Lifshitz solutions.

The variation of the action about a solution of the equations of
motion is just the boundary term,
\begin{eqnarray}
\delta I&=&\frac{1}{16\pi}\int_{\partial\mathcal{M}}d^{n}x\sqrt{-h}%
\{\Pi_{\alpha\beta}\delta
h^{\alpha\beta}-(1-2\omega)n^{\mu}F_{\mu\nu}\delta
A^{\nu}-n^{\mu}H_{\mu\nu}\delta B^{\nu} \\
&& +f^{\prime}(B_{\alpha}B^{\alpha})(2B_{\alpha}\delta
B^{\alpha}+B_{\alpha}B_{\beta}\delta h^{\alpha\beta})-\frac{1}{2}%
f(B_{\alpha}B^{\alpha})h_{\alpha\beta}\delta h^{\alpha\beta}\},
\nonumber
\end{eqnarray}
where
\begin{equation}
\Pi_{\alpha\beta}=K_{\alpha\beta}-Kh_{\alpha\beta}+(n-1)h_{\alpha\beta}.
\end{equation}
Now, if one defines
\begin{eqnarray}
&&S_{\alpha \beta }=\frac{\sqrt{-h}}{16\pi }\left[ \Pi _{\alpha \beta }+%
\frac{zq}{2}(-B_{\gamma }B^{\gamma })^{-1/2}(B_{\alpha }B_{\beta
}-B_{\gamma
}B^{\gamma }h_{\alpha \beta })\right] , \\
&&S_{\alpha }^{L}=-\frac{\sqrt{-h}}{16\pi }\left[ n^{\mu }H_{\mu
\alpha
}-zq(-B_{\gamma }B^{\gamma })^{-1/2}B_{\alpha }\right] , \\
&&S_{\alpha }^{M}=-\frac{\sqrt{-h}}{16\pi }(1-2\omega)n^{\mu
}F_{\mu \alpha },
\end{eqnarray}
Then the general variation of the action is\footnote{As we shall see, it is only in the
canonical ensemble that subsequent equations in this section are defined. Note that $\delta
A^{\alpha } = 0$ in the grand canonical ensemble.}
\begin{equation}\label{delta-I}
\delta I=\int d^{n}x(S_{\alpha \beta }\delta h^{\alpha \beta
}+S_{\alpha }^{L}\delta B^{\alpha }+S_{\alpha }^{M}\delta
A^{\alpha })
\end{equation}

Accordingly we can define a stress tensor complex \cite{Saremi}
consisting of the energy density $\mathcal{E}$, energy flux
$\mathcal{E}_{i}$, momentum density $\mathcal{P}_{i}$ and spatial
stress tensor $\Pi _{ij}$, satisfying the conservation equations
\begin{equation}
\partial _{t}\mathcal{E}+\partial _{i}\mathcal{E}^{i}=0,\qquad \partial _{t}%
\mathcal{P}_{j}+\partial _{i}\mathcal{P}_{\phantom{i}{j}}^{i}=0,
\end{equation}
where
\begin{equation}
\mathcal{E}=2S_{\phantom{t}{t}}^{t}-S_{L}^{t}B_{t}-S_{M}^{t}A_{t},\qquad \mathcal{E}^{i}=2S_{%
\phantom{i}{t}}^{i}-S_{L}^{i}B_{t}-S_{M}^{i}A_{t},
\end{equation}
and
\begin{equation}
\mathcal{P}_{i}=-2S_{\phantom{t}{i}}^{t}+S_{L}^{t}B_{i}+S_{M}^{t}A_{i},\qquad \mathcal{P}_{%
\phantom{j}{i}}^{j}=-2S_{\phantom{j}{i}}^{j}+S_{L}^{j}B_{i}+S_{M}^{j}A_{i}.
\end{equation}
Using the exponential ansatz (\ref{newansatz}) for the metric and
gauge potentials,we obtain
\begin{equation}
\mathcal{E}=\frac{1}{16\pi } \left[ r^{n+z-1}\frac{f}{g}[%
zq^{2}(1-h)jg-2(n-1)(1-g)]+(1-2\omega)Qr^z\kappa
\right]_{r\rightarrow \infty }.\label{Energy}
\end{equation}
for the energy density of the black brane.

Employing the large $r$ expansions given in Appendix for the
metric functions in the case of $z\neq n-1$ and $z=n-1$ to eq.
(\ref{Energy}), we obtain the rather curious result that
\begin{equation}
\mathcal{E}=\left\{
\begin{array}{cc}
\frac{1}{8\pi
}\frac{(z-1)(n-z-1)}{(n-1)}C_{1}+\frac{(4\omega-3) Q^2}{32\pi(n-z-1)}r^{z-(n-1)}& \, \, z\neq n-1,\, z\neq1\\ & \\
\frac{1}{16\pi(n-1)}C_1+
\frac{Q^2}{32\pi(n-1)}[(3-4\omega)(n-1)\ln r +1]& \,\,z=n-1
\end{array}
\right. \label{energy}
\end{equation}
giving a finite energy density  if $z<n-1$
regardless of the choice of  $\omega$.  We remark that for arbitrary $z$
choosing $\omega=\frac{3}{4}$ yields finiteness of energy.  However it is straightforward to
show that  $S_{\alpha }^{M}$ is divergent for large $r$,  rendering the variation (\ref{delta-I}) ill-defined,
unless $\omega = 1/2$.

Choosing $\omega= \frac{1}{2}$ gives
\begin{eqnarray}
I&=&\frac{1}{16\pi}\int_{\mathcal{M}}d^{n+1}x\sqrt{-g}(\mathcal{L}_{g}+%
\mathcal{L}_{m}) \\
&&+\frac{1}{8\pi}\int_{\partial\mathcal{M}}d^{n}xz\sqrt{-h}\left(K-(n-1)-%
\frac{zq}{2}\sqrt{-B_{\alpha}B^{\alpha}}+\frac{1}{2}
n^{\mu}F_{\mu\nu} A^{\nu}\right)
\end{eqnarray}
for the action.
Using equations (\ref{Ceqn1}-\ref{Chorizon}) we obtain
\begin{equation}
C_1=\frac{8\pi(n-1)^2}{(z-1)(n+z-1)(n-z-1)}\left[ TS+D\mathcal{Q}_{}\right]
\qquad z< n-1,
 \label{C1}
\end{equation}
and applying (\ref{C1}) we find
\begin{equation}
\mathcal{E}=\frac{n-1}{n+z-1}\left[TS+\mathcal{Q}D\right] \quad  z< n-1,
 \label{ene}
\end{equation}
Since (\ref{ene}) is the expression for energy at infinity one may
interpret the coefficient conjugate to the electric charge as the
chemical potential. Hence
\begin{equation}
\Phi=D  \quad  z< n-1,
\label{Chempot}
\end{equation}
which is consistent with the definition of chemical potential, measured
at infinity with respect to the horizon, given as \cite{Cal}
\begin{eqnarray}
\Phi &=&A_{\mu }\chi ^{\mu }\left| _{r\rightarrow \infty }-A_{\mu
}\chi ^{\mu }\right| _{r=r_{0}} = Q\int^{\infty} \frac{fg}{r^{n-z}}dr+D \nonumber\\
&=& D \quad  z< n-1 \label{Pot3}
\end{eqnarray}
where $\chi^{\mu}=\partial _{t}$ is the null generator of the
horizon.
For $z=1$ setting $q=0$, substituting (\ref{frads}) in eq. (\ref{Energy}),
and expanding the result for large $r$ yields
the energy density of Reisner-Nordstrom black brane
\cite{DehKhod}:
\begin{equation}
\mathcal{E}=\frac{(n-1)}{16\pi}m,
\end{equation}
where $m=\mathcal{C}/n$ is the mass of the AdS black hole. The quantity
$\mathcal{C}$ is the conserved charge given by (\ref{consads}),
which at the horizon in terms of thermodynamic quantities and
horizon radius can be expressed as
\begin{equation}
\mathcal{C}=16\pi
\left[TS+\mathcal{Q}\frac{Q}{(n-2)r_0^{n-2}}\right] \quad\quad z=1
\end{equation}
and the coefficient of electric charge is exactly AdS chemical
potential \cite{DehKhod}:
\begin{equation}
\Phi=D=-Q\int^{r_0} \frac{dr}{r^{n-1}}=\frac{Q}{(n-2)r_0^{n-2}} \quad\quad z=1 \label{chempotads}
\end{equation}
Therefore, the energy density for $z=1$ is
\begin{equation}
\mathcal{E}=\frac{n-1}{n}(TS+\mathcal{Q}\Phi).\label{eneads}
\end{equation}
which, by comparing Eqs. (\ref{ene}), (\ref{Chempot}) and (\ref{eneads}), we see generalizes to
\begin{equation}
\mathcal{E}=\frac{n-1}{n+z-1}(TS+\mathcal{Q}\Phi) , \label{eneLif}
\end{equation}
for  asymptotic Lifshitz black branes.

\section{Concluding Remarks}

Our considerations of Lifshitz gravity coupled to electromagnetism
in $(n+1)$ dimensions have indicated  a rich array of
numerical solutions that depend on the two parameters $h_0$ and $Q$. Solutions exist for
all topologies, and an extremal limit, denoted by $Q_c$ and
defined in eq (\ref{Qc}) exists.

The general effect of $Q$ is to cause metric functions to more slowly approach their asymptotic values,
other parameters being equal.  We illustrated this with the metric function $f(r)$: it  tends to grow more slowly toward its asymptotic value with increasing charge (as shown in section 6).  We also found that increasing charge decreases the boundary length of the Wilson loop (for $n=3$) while causing the potential $V$ between two particles to grow more rapidly with increasing
string midpoint length $r_m$.

We also found that we can extend  the thermodynamics of Lifshitz black branes \cite{Peet,Peet2,HR} to the charged case, provided
the parameter $z<n-1$. We have obtained an expression, eq. (\ref{eneLif}), for the energy density in terms of the extensive thermodynamic  quantities entropy and charge density and their intensive conjugate quantities for asymptotic Lifshitz black branes, generalizing the $z=1$ AdS case (\ref{eneads}).  Extending to values of $z$ larger than this entails a choice of boundary terms that renders the variational principle ill-defined.   Commensurate results have been obtained in other models \cite{Peet3} whose solutions smoothly interpolate between Lifshitz-like and AdS-like behaviour as the ratio of $T/\mu$ (with $\mu$ the chemical potential)
varies from small to large values.

Further work in this area will involve obtaining a better understanding of how the general effects in this paper affect the dual
theories associated with asymptotically Lifshitz spacetimes, and of how quantum-gravitational corrections can likewise modify such effects.

\section*{Acknowledgements}
This work was supported in part by the Natural Sciences and Engineering Research Council of Canada.

\section*{Appendix}

Here we are willing to explore more details of the eigen modes at
large $r$ which have been roughly introduced in eq. (\ref{eigen}).
Here we ignore the universal mode in the set ({\ref{Eqp}}) by
setting $k=0$.  The complete solution to ({\ref{Eqp}}) at large
$r$ provided $z\neq n-1$ is
\begin{eqnarray}
&&g_1(r)=-\frac{C_{1}G_{1}}{r^{z+n-1}}-\frac{C_{2}G_{2}}{r^{(z+n-1+\sqrt{%
\gamma})/2}}-\frac{C_{3}G_{3}}{r^{(z+n-1-\sqrt{\gamma})/2}}-\frac{(n-2z)Q^2}{4(n+z-2)(n-z-1)^2r^{2n-2}},  \label{glr} \\
&&h_1(r)=-\frac{C_{1}}{r^{z+n-1}}-\frac{C_{2}}{r^{(z+n-1+\sqrt{\gamma})/2}}%
-\frac{C_{3}}{r^{(z+n-1-\sqrt{\gamma})/2}}+\frac{(2n-z-2)Q^2}{4(n+z-2)(n-z-1)^2r^{2n-2}},  \label{hlr} \\
&&j_1(r)=-\frac{C_{1}J_{1}}{r^{z+n-1}}-\frac{C_{2}J_{2}}{r^{(z+n-1+\sqrt{%
\gamma})/2}}-\frac{C_{3}J_{3}}{r^{(z+n-1-\sqrt{\gamma})/2}}-\frac{Q^2}{4(n-z-1)^2r^{2n-2}},
\label{jlr}
\end{eqnarray}
where
\begin{eqnarray}
&&\gamma=9z^{2}-2(3n+1)z+(n^{2}+6n-7), \\
&&G_{1}=\frac{z(z-1)}{(n-1)^2},\quad G_{2}=\frac{z-1}{n-1},\quad G_{3}=\frac{%
z-1}{n-1}, \\
&&J_{1}=-{\frac{z(z-2+n)}{(n-1)^{2}}},\quad J_{2}=\frac{n-3z+1-\sqrt{\gamma}%
}{2(n-1)},\quad J_{3}=\frac{n-3z+1+\sqrt{\gamma}}{2(n-1)}.
\end{eqnarray}
If one inserts the above expression in equation for the small
perturbation of $f(r)$, i.e
\begin{equation}
r\frac{d}{dr}f_1(r)=(n-2+2z)g_1(r)-\frac{z(z-1)}{n-1}
h_1(r)+(z-1)j_1(r)-\frac{Q^2}{4(n-1)r^{2n-2}}+\frac{(n-2)k}{2r^{2}},
\end{equation}
then again the universal $1/r^2$ mode decays away at infinity and
one obtains
\begin{equation}
f_1(r)=\frac{C_{1}F_{1}}{r^{z+n-1}}+\frac{C_{2}F_{2}}{r^{(z+n-1+\sqrt{%
\gamma})/2}}+\frac{C_{3}F_{3}}{r^{(z+n-1-\sqrt{\gamma})/2}}+\frac{Q^2}{4(n-z-1)(n-1)r^{2n-2}},
\label{flr}
\end{equation}
with the constraint that $f_1(r)$ should goes to zero as r goes to
infinity and
\begin{eqnarray}
&&F_{1}=\frac{z(z-1)(n-z-1)}{(n-1)^{2}(n+z-1)},\quad
F_{2}=\frac{z(n+1)-n^2-2z^2+1+(n-1)\sqrt{\gamma}}{2(n-1)(n-z-1)}, \nonumber\\
&&F_{3}=\frac{z(n+1)-n^2-2z^2+1-(n-1)\sqrt{\gamma}}{2(n-1)(n-z-1)}.
\end{eqnarray}
For $z=n-1$ solutions are given by:
\begin{eqnarray}
g_1(r)&=&-\frac{(n-2)}{2(n-1)^2}\frac{(C_{1}\ln{r}+C_{2})}{r^{2n-2}}
-\frac{(3n-4)}{4(n-1)^3}\frac{C_1}{r^{2n-2}}\nonumber\\
&+&\left[(n-2)(\ln{r})^2+\frac{(3n-4)\ln{r}}{n-1}+\frac{5n-6}{2(n-1)^2}\right]\frac{Q^2}{8(n-1)r^{2n-2}}, \\
h_1(r)&=&-\frac{(C_{1}\ln{r}+C_{2})}{2(n-1)r^{2n-2}}%
-\frac{C_{1}}{4(n-1)^2r^{2n-2}}+\left[(\ln{r})^2+\frac{\ln{r}}{n-1}+\frac{1}{2(n-1)^2}\right]\frac{Q^2}{8r^{2n-2}}, \\
j_1(r)&=&\frac{(2n-3)}{2(n-1)^2}\frac{(C_{1}\ln{r}+C_{2})}{r^{2n-2}}%
+\frac{(2n-3)}{4(n-1)^{3}}\frac{C_{1}}{r^{2n-2}}\nonumber\\
&-&\left[(2n-3)(\ln{r})^2+\frac{(2n-3)\ln{r}}{n-1}+\frac{4n-5}{2(n-1)^2}\right]\frac{Q^2}{8(n-1)r^{2n-2}}, \\
f_1(r)&=&\frac{(3n-4)}{4(n-1)^3}\frac{C_{1}}{r^{2n-2}}-\left[(n-1)\ln{r}+1\right]\frac{(3n-4)Q^2}{8(n-1)^3r^{2n-2}}.
\label{f1rz}
\end{eqnarray}
which are relevant to asymptotic Lifshitz background.

\end{document}